\documentclass[manuscript]{aastex62}

\usepackage{graphicx}
\usepackage{placeins}
\usepackage{morefloats}

\shorttitle{THE RELATIVE EMISSION FROM CHROMOSPHERES AND CORONAE} 
\shortauthors{Linsky, Wood, Youngblood, Brown, et al.}

\begin{document}
\title{THE RELATIVE EMISSION FROM CHROMOSPHERES AND CORONAE: DEPENDENCE ON
  SPECTRAL TYPE AND AGE\footnote{Based on observations made with the NASA/ESA
Hubble Space Telescope obtained from the Mikulski Archive for Space
Telescopes (MAST) at the Space
Telescope Science Institute, which is operated by the Association of
Universities for Research in Astronomy, Inc., under NASA contract NAS
AR-09525.01A. These observations are associated with programs \#12475,
12596, 13650, 14640, and 15071.}}

v

\author[0000-0003-4446-3181]{Jeffrey L. Linsky}
\affiliation{JILA, University of Colorado and NIST, Boulder, CO 
80309-0440, USA}

\author[0000-0002-4998-0893]{Brian E. Wood}
\affiliation{Naval Research Laboratory, Space Sciences Division, 
  Washington, DC 20375, USA}

\author{Allison Youngblood}
\affiliation{Laboratory for Atmospheric and Space Physics, University
  of Colorado, 600 UCB, Boulder, CO 80309-0600, USA}
 
\author{Alexander Brown}
\affiliation{CASA, University of Colorado, Boulder, CO 80309, USA}

\author{Cynthia S. Froning}
\affiliation{McDonald Observatory, University of Texas at Austin, Austin, TX
  78712, USA}

\author{Kevin France}
\affiliation{Laboratory for Atmospheric and Space Physics, University of
  Colorado, 600 UCB, Boulder, CO 80309-0600, USA}
\affiliation{Department of Astrophysical and Planetary Sciences, University of
  Colorado, Boulder, CO 80309, USA}

\author{Andrea P. Buccino}
\affiliation{Dpto. de Fisica, Facultad de Ciencias Exactas y Naturales (FCEN)
  Universidad de Buenos Aires (UBA), Buenos Aires, Argentina}
\affiliation{Instituto de Astronomia y Fisica del Espacio (CONICET-UBA),
   Buenos Aires, Argentina}

\author[0000-0002-3699-3134]{Steven R. Cranmer}
\affiliation{Laboratory for Atmospheric and Space Physics, University of
  Colorado, 600 UCB, Boulder, CO 80309-0600, USA}
\affiliation{Department of Astrophysical and Planetary Sciences, University of
  Colorado, Boulder, CO 80309, USA}

\author{Pablo Mauas}
\affiliation{Dpto. de Fisica, Facultad de Ciencias Exactas y Naturales (FCEN)
  Universidad de Buenos Aires (UBA), Buenos Aires, Argentina}
\affiliation{Instituto de Astronomia y Fisica del Espacio (CONICET-UBA),
   Buenos Aires, Argentina}

\author{Yamila Miguel}
\affiliation{Leiden Observatory, P.O. Box 9500, 2300 RA Leiden, The Netherlands}

\author[0000-0002-4489-0135]{J. Sebastian Pineda}
\affiliation{Laboratory for Atmospheric and Space Physics, University
  of Colorado, 600 UCB, Boulder, CO 80309-0600, USA}

\author{Sarah Rugheimer}
\affiliation{University of Oxford, Clarenden Laboratory, AOPP, Sherrington
  Road, Oxford, OX1 3PU, UK}

\author{Mariela Vieytes}
\affiliation{Instituto de Astronomia y Fisica del Espacio (CONICET-UBA),
   Buenos Aires, Argentina}

\author{Peter J. Wheatley}
\affiliation{Department of Physics, University of Warwick, Coventry 
   CV4 7AL, UK}

\author{David J. Wilson}
\affiliation{McDonald Observatory, University of Texas at Austin, Austin, TX
  78712, USA}

\correspondingauthor{Jeffrey L. Linsky}
\email{jlinsky@jila.colorado.edu}



\today

\begin{abstract}
Extreme-ultraviolet and X-ray emissions from stellar coronae drive mass loss
from exoplanet atmospheres, and ultraviolet emission from stellar
chromospheres drives
photo-chemistry in exoplanet atmospheres. Comparisons of the spectral
energy distributions of host stars are, therefore, essential for
understanding the evolution and habitability of exoplanets.
The large number of stars observed with the MUSCLES, Mega-MUSCLES, and other
recent {\em HST} observing programs has provided for the first time
a large sample (79 stars) of
reconstructed Ly-$\alpha$ fluxes that we compare with X-ray fluxes to 
identify significant patterns in the relative
emission from these two atmospheric regions as a function of stellar age and
effective temperature.   
We find that as stars age on the main sequence, the emissions from their
chromospheres and coronae
follow a pattern in response to the amount of magnetic heating in these
atmospheric layers. A single trendline slope describes the pattern
of X-ray vs. Lyman-$\alpha$ emission for G and K dwarfs, but the different
trendlines for M dwarf stars show that the Ly-$\alpha$
fluxes of M stars are significantly smaller than warmer stars with the
same X-ray flux.
The X-ray and Lyman-$\alpha$ luminosities divided by the stellar
bolometric luminosities show different patterns depending on stellar age.
The L(Ly-$\alpha$)/L(bol) ratios increase smoothly to cooler stars of all ages,
but the L(X)/L(bol) ratios show different trends.
For older stars, the increase in coronal emission
with decreasing $T_{\rm eff}$ is much steeper 
than chromospheric emission. We suggest a fundamental
link between atmospheric properties and trendlines relating coronal and
chromospheric heating,

\end{abstract}


\section{Introduction}

     Essentially all stars with convective interiors from the A7~V star
$\alpha$~Aql \citep{Robrade2009} to the late-M and perhaps L dwarfs 
\citep{Berger2010,Stelzer2012,Hawley2003} 
emit ultraviolet (91.2--300~nm, UV) and X-ray (0.1--10~nm) photons
from plasmas at temperatures 
 ranging from roughly 5,000~K to at least $10^6$~K. Since these
plasmas are too hot to be explained by radiative/convective equilibrium in
stellar photospheres, there must be an additional heat source to explain
their elevated temperatures. This heat source is either
the dissipation of MHD waves or direct magnetic field reconnection events
called flaring, see review by \cite{Cranmer2019}. 
In analogy with the solar chromosphere and corona, stellar plasmas
in the lower temperature range are called chromospheres 
\citep[see review by][]{Linsky2017} and plasmas in the higher temperature 
range are called coronae \citep[see review by][]{Gudel2004}.
Except for very faint stars where instrumental sensitivity limits detection,
all dwarf stars with convective interiors show X-ray emission from
coronae and emission lines of Mg~II, Ca~II, and H~I Lyman-$\alpha$ indicating
the presence of plasma at chromospheric temperatures.

There are many examples of correlations of chromospheric
emission (e.g., Lyman-$\alpha$, Ca~II H and K lines, Mg~II h and k
lines, H$\alpha$) with such stellar activity indicators as age,
rotation, and magnetic field strength and coverage 
\citep[e.g.,][]{Wood2005,Guinan2016,Newton2017}. There are also
correlations of coronal properties such as X-ray emission with
activity indicators \citep[e.g.,][]{Gudel2004,Wood2005}.
These correlations generally show saturation at
high activity levels and linear regressions in log-log plots with
decreasing activity indicators
such as age and rotation. These correlations are usually
described by power-law relations of the form
$\log F$(corona) $= \alpha \log F$(chromo) $+\beta$, 
where $F$(corona) is a coronal flux
or luminosity diagnostic, usually the broad band X-ray emission, and
$F$(chromo) is the flux or luminosity of a chromospheric diagnostic,
generally an emission line such as the Ca~II K line (393.3~nm), Mg~II k line
(279.6~nm) or H~I Lyman-$\alpha$ line (121.56~nm). An example of such
power-law correlations between X-ray and Mg~II emission 
is $\alpha=2.20\pm 0.13$ for F and G dwarfs and
$\alpha=2.90\pm 0.20$ for K dwarfs \citep{Wood2005}. Another example
is the correlations of chromospheric Ca~II K line with
UV emission lines and extreme-ultraviolet (10--91~nm, EUV) flux
\citep{Youngblood2017}.

The correlations have  steeper slopes for activity indicators formed at
higher temperatures in the stellar atmosphere. For example, in a volume-limited
sample of 159 M dwarfs located within 10~pc, \cite{Stelzer2013} found 
that the slope of X-ray lumiosity with age is steeper than that for the
luminosity in the {\em Galaxy Evolution Explorer (GALEX)}
far-ultraviolet (134--178~nm, FUV) emission formed in the upper chromosphere
and the {\em GALEX} near-ultraviolet (177--283~nm, NUV)
emission formed in the lower chromosphere.
\cite{Ribas2005} found a steeper decline of X-ray emission with age
compared to chromospheric and transition region emission for solar analog
stars, and \cite{Guinan2016} found a steeper decline of X-ray emission
compared to Ly-$\alpha$ emission for M0-M5~V stars. For M stars,
\cite{Guinan2016} showed that the decay
of X-ray emission with time is also faster than the decay of
Lyman-$\alpha$ radiation. These pioneering studies point to a trend
of decreasing stellar activity that results from 
the decay of magnetic fields with decreasing rotation rate. 
The age dependence of decreasing activity depends on stellar mass with
the decay time scale of order 100 Myr for F--K dwarfs and increasing to 
several Gyr for late-M dwarfs \citep{Reiners-Mohanty2012}.

Flux-flux diagrams, which plot one emission feature, such as X-rays
vs. another emission feature such as UV emission or Ca~II emission,
are useful tools for studying the spectral energy distributions of
radiation seen by exoplanets. \cite{Stelzer2013}
found that in plots of X-ray luminosity vs. luminosity in the {\em GALEX}
FUV  and NUV  bands, M dwarfs with 
weak emission follow the same trendline as
active M dwarfs over three orders of magnitude in X-ray luminosity.
\cite{Walkowicz2009} showed that the
X-ray flux and Ca~II emission follow a similar trendline for M3~V stars.

\cite{Oranje1986,Schrijver1987} and \cite{Rutten1989} observed 
chromospheric emission in the Mg~II lines 
with the {\it International
Ultraviolet Explorer (IUE)} satellite and Ca~II emission from
ground-based observatories together with X-ray emission observed by 
the {\it European X-ray Observatory Satellite (EXOSAT)} for main sequence
stars. They found that M dwarfs deviate
from the chromosphere-coronal flux-flux correlation (here called
trendlines) seen in the warmer stars in
the sense that the chromospheric emission is systematically weak
compared to coronal emission and that this weakness becomes more pronounced
for the coolest and least active stars. 

Increasing interest in the environments and habitability of 
exoplanets of M dwarfs \citep{Shields2016,Kaltenegger2017,Wandel2018}
and the availability of more sensitive
ultraviolet spectra with the {\it Hubble Space Telescope (HST)}
and X-ray fluxes with the {\it R\"ontgensatellit (ROSAT), Chandra,} and
{\it XMM-Newton} satellites encouraged us to re-examine the difference in the
trendlines between M dwarfs and warmer
stars. In particular, we explore whether there is a difference between
the trendlines of warmer and cooler M dwarfs and whether 
stellar age and rotation set
limits for these trendlines. Given that the emission from  M dwarfs is variable
at all wavelengths, our study benefits from the
availability of near simultaneous high-resolution UV spectra of M
stars obtained for some of the stars in the 
{\it HST} MUSCLES Treasury Survey program
\citep{France2016}. We will also use UV spectra obtained with other
{\it HST} programs specifically aimed at M dwarfs. 

This paper will explore the X-ray
vs. Lyman-$\alpha$ trendlines for M dwarfs separated into spectral
type and age bins. X-ray emission is the primary tool for measuring the
heating rates in stellar coronae, although thermal conduction, winds,
and radiation in the EUV are additional sinks for coronal heating. The
Lyman-$\alpha$ emission line is by far the brightest feature emitted
by chromospheres of G-type stars and represents at least half of the
total emission in the ultraviolet spectra of M dwarfs \citep{France2012}.
\cite{Claire2012} estimated that Ly-$\alpha$ photons represented about 40\%
of all solar photons at $\lambda < 170$~nm throughout the Sun's history.
Compared to other chromospheric emission lines formed at temperatures
less than 10,000~K, the power-law index $a$ in the relation
log F(Ly-$\alpha$) = a log F(line) + b is close to unity:
$a=0.77\pm 0.11$ for the Mg~II k line \citep{Youngblood2016}
  and $a=0.88\pm 0.11$ for the Ca~II K line \citep{Youngblood2017}. Thus the
  flux in the Ly-$\alpha$ line is a good but not perfect proxy for the total
  emission from a stellar chromosphere at temperatures less than about
  15,000~K.
In our comparison of coronal and chromospheric emission from stars
between spectral types F and late-M, we will use X-ray emission as a
diagnostic of coronal emission and Lyman-$\alpha$ fluxes reconstructed
to remove interstellar absorption
as a diagnostic for chromospheric emission.

In Section 2, we list the available reconstructed Lyman-$\alpha$ and
X-ray data and the origins of these data. Section~3 describes the different
trendlines for the warmer stars compared to the M dwarfs, and in
Section~4 we compare L(X)/L(bol) with L(Ly$\alpha$)/L(bol) as functions of
stellar effective temperature and age.
We then consider possible explanations for the different
trendline slopes in Section~5, and summarize our conclusions in Section~6.   

\section{Lyman-$\alpha$ and X-ray fluxes for F-M dwarf stars}

We include in this study all dwarf stars with spectral types later
than mid-F for which there are both
reconstructed Lyman-$\alpha$ and broad band X-ray fluxes. A major
source of these data is the paper by \cite{Linsky2013} that gives
Lyman-$\alpha$ and X-ray fluxes for five F stars, 18 G stars, 16 K
stars, and 8 M stars. The coolest M dwarf in this list is 
Proxima Centauri (M5.5~V). 
The Lyman-$\alpha$ fluxes were reconstructed either by correcting
for the observed interstellar H~I Lyman-$\alpha$ absorption along the lines of
sight to the stars \citep{Wood2005}, or by
simultaneously solving for the intrinsic line profile and the
interstellar absorption \citep{France2012,Youngblood2016,Wilson2020}. 
The observations and their data sources
are listed in Table~1. The Lyman-$\alpha$ and
X-ray fluxes (erg~cm$^{-2}$~s$^{-1}$) are listed at a standard 
distance of 1 au. The stellar effective temperatures and ages
are primarily from \cite{Schneider2019} and \cite{Melbourne2020}.
The bolometric luminosities are
computed from the effective temperatures, stellar radii, and {\em GAIA}
parallaxes cited in these papers. For many of the stars these quantities
are from Youngblood (in prep.). 
In addition to the Ly-$\alpha$ and X-ray fluxes cited in the \cite{Linsky2013}
paper, we include new data from the following sources:

\begin{description}

\item[MUSCLES Treasury Survey] The {\em Measurements of the Ultraviolet
  Spectral Characteristics of Low-mass Exoplanetary Systems (MUSCLES)}
  Treasury Survey \citep{France2016, Youngblood2016, Loyd2016}
  observed seven low activity M dwarfs and 4 K dwarfs
  together with coordinated X-ray and ground-based observations.  We
  include in Table~1 the reconstructed Lyman-$\alpha$ fluxes
  \citep{Youngblood2016} and
  coordinated X-ray fluxes corrected for interstellar absorption
  \citep{Loyd2016}. The {\em MUSCLES} fluxes for the stars GJ~667C, GJ 832, 
  GJ 876, GJ 581, and GJ 436 are listed in Table~1 instead of the
  values listed in the previous \cite{Linsky2013} paper. Proxima Centauri
  was not part of the MUSCLES survey, but the fluxes for the 2017
  {\em HST} and {\em Chandra} observations of the star are included in
  the MUSCLES Team webpage.

\item[Mega-MUSCLES Program] Continuing from the MUSCLES program,
  the {\em Mega-MUSCLES} program
  \citep{Froning2019,Wilson2020} observed 13 more active M dwarfs with {\em HST}
  (program GO-15071) with coordinated {\em Chandra,  XMM-Newton} and
  ground-based observations. \cite{Youngblood2020} have reconstructed 
  the Lyman-$\alpha$ fluxes for these stars using the technique described in
  \cite{Youngblood2016}. Brown et al. (in prep) have analyzed the
  {\em Chandra} ACIS-S S3 CCD spectra of five of these stars: GJ~15A, GJ~163,
  GJ~849, LHS~2686, and GJ~699 (Barnard's star).
  The fluxes listed in Table~1 are mean values,
  except that a flare on GJ~799 was deleted to provide the quiescent emission
  level. The {\em Chandra} observation of GJ~15A was simultaneous with the
  {\em HST} observation.\\

  One of the Mega-MUSCLES targets, the M7.5~V star
  TRAPPIST-1 has a previous reconstructed Lyman-$\alpha$ flux
  \citep{Bourrier2017} and X-ray flux \citep{Wheatley2017}. We use the
  new Mega-MUSCLES fluxes \citep{Wilson2020} because the {\em HST} and
  {\em XMM-Newton} observations are nearly simultaneous and thus can be
  reliably intercompared. To obtain a rough estimate of the uncertainty in
  the Lyman-$\alpha$ flux of a late-M dwarf like TRAPPIST-1, one should consider
  both possible errors in the reconstruction technique
  and stellar variability. \cite{Bourrier2017}
  obtained a reconstructed Lyman-$\alpha$ flux of
  $7.6^{+1.5}_{-3.0}\times 10^{-15}$ erg~cm$^{-2}$~s$^{-1}$ from their 2016
  STIS data, but using
  a different reconstruction technique \cite{Wilson2020} obtained
  $(1.09^{+0.40}_{-0.27})\times 10^{-14}$ when analyzing the same data.
  The analysis of
  the 2018 Ly-$\alpha$ STIS spectra by \cite{Wilson2020} resulted in a
  reconstructed flux of $(1.40^{+0.60}_{-0.36})\times 10^{-14}$ in the same
  units. The uncertainty in the X-ray flux of TRAPPIST-1 is primarily due to
  stellar variability. \cite{Wheatley2017} obtained an
  X-ray flux of $(2.0-4.3)\times 10^{-14}$ erg~cm$^{-2}$~s$^{-1}$ from their
  2014 {\em XMM-Newton} EPIC observation, while \cite{Wilson2020} obtained
  $2\times 10^{-14}$ from their 2018 EPIC observation.
  
\item[High radial velocity stars program] We include the reconstructed
  Lyman-$\alpha$ line fluxes for Kapteyn's star
  \citep{Guinan2016,Youngblood2016} and 
  the reconstructed Lyman-$\alpha$ fluxes for two high-radial
  velocity stars with spectral types G8~V to M4~V observed by 
  Youngblood et al. (in prep.) obtained with {\em HST} program GO-15190. We also
  include the high radial velocity stars Ross~825 and Ross~1044
  analyzed by \cite{Schneider2019}.

\item[Winds of M dwarf stars program] We include the reconstructed 
  Lyman-$\alpha$ and X-ray fluxes of nine nearby M dwarf stars observed
  by Wood et al. (in prep.) with {\em HST} program GO-15326. These stars
  were observed primarily to measure
  their winds using the Lyman-$\alpha$ line.

\item[FUMES targets] The {\em Far-Ultraviolet M-dwarf Evolution Survey
  (FUMES)} (Pineda et al. in prep.) observed ten M0 V to M5~V stars. We
  have included the two stars with STIS E-140M moderate resolution
  spectra and three of the eight stars with STIS G-140L low resolution
  spectra. Youngblood et al. (in prep.) reconstructed the Lyman-$\alpha$
  lines of these five stars with cited measurement uncertainties less 
   than 25\%. 

\item[Individual targets] We also include in Table~1 the reconstructed
  Lyman-$\alpha$ line fluxes of the stars: HD~28568 \citep{Schneider2019},  
  $\pi$~Men \citep{GarciaMunoz2020}, 
  Kepler-444 \citep{Bourrier2017}, 
  Kapteyn's star \citep{Guinan2016}, GJ~3470 \citep{Bourrier2018}, 
  GJ~821 and GJ~213 \citep{Youngblood2017}, GJ~1132 \citep{Waalkes2019},
  and TRAPPIST-1 \citep{Wheatley2017,Bourrier2017}.

\end{description}

\subsection{X-ray Fluxes}

The X-ray fluxes cited in Table~1 were observed by instruments on
{\em XMM-Newton, Chandra}, and {\em ROSAT}. Fluxes are given for the
0.2--12~keV (0.1--6~nm) band. For the MUSCLES and Mega-MUSCLES surveys,
some of the X-ray observations were obtained on the same day or
adjacent days as the Ly-$\alpha$ observations, but most of the X-ray
observations in Table~1 were obtained at random times
uncorrelated with the Ly-$\alpha$ observations. We have identified two stars
in the MUSCLES program where all of the X-ray and Ly-$\alpha$ observations
were obtained within the same day. In the subsequent figures, these two M dwarfs
(GJ~667C and GJ~176) are identified with circuled symbols.

A major source of X-ray fluxes is from the on-line {\em XMM}
Serendipitous Source
Catalog\footnote{https://heasarc.gsfc.nasa.gov/W3Browse/xmm-newton/xmmssc.html}. The fourth
generation catalog (4XMM-DR9), which is a complete re-reduction of all
data obtained by {\em XMM-Newton} until March 2019, 
contains the mean fluxes from pointed observations, which may be many pointings
for a given star. Another source of X-ray data is the {XMM-Newton} Slew Source
Catalog XMMSL2\footnote{https://www.cosmos.esa.int/web/xmm-newton/xmmsl2-ug}.
A description of the first slew catalog is in \cite{Saxton2008}.
From sources in both catalogs, we selected data with the smallest error bars.

\section{Stellar Flux-Flux Trendlines}

   Figure~1 plots the reconstructed Lyman-$\alpha$ and X-ray fluxes 
listed in Table~1 for the F, G,
and K stars. X-ray and Lyman-$\alpha$ fluxes at the standard distance
of 1 au are plotted with different symbols and colors for each
spectral class. The least-squares linear fits to the data sets of the 
F, G and K stars all have the same slopes, and the trendlines for the G and K
stars overlap. We note that the most active single star Speedy Mic (K3~V)
  lies at the top of the K star trendline next to the spectroscopic binary
  V471~Tau (K2~V + WD), and that the least active K star as measured by the
  weakest X-ray flux, HD~40307 (K2.5~V), lies below the K star trendline.
  In this and subsequent figures, we do not plot
  measurement errors because a prime source of error is stellar activity, which
  is different for the UV and X-ray data typically obtained at different times
  especially for cooler stars.

Figure 2 shows the same data as Figure~1 but includes the M dwarfs
divided into M0~V -- M2.5~V, and M3~V -- M7.5~V groups.
The early-M star with the weakest
Lyman-$\alpha$ flux is the subdwarf Kapteyn's star (M1~VI). It's weak
Lyman-$\alpha$ emission may result from the star's low metallicity and thus low
electron density at chromospheric temperatures.

We plot linear least-squares fits to the data for each spectral type
group. The fits to the F2--G9 and K stars are similar with $\alpha$
equal to $2.32\pm 0.26$, and $2.34\pm 0.31$, respectively.
These values are similar to the ones reported by \cite{Wood2005}:
$\alpha=2.20\pm 0.13$ for the F-G dwarfs and $\alpha=2.90\pm 0.20$
for the K dwarfs, as would be expected since there are many
stars in common among the two data sets. A single least
squares fit with $\alpha \approx 2.3$ would fit all of the G and 
K stars including the Sun observed at low activity.

     The M stars, however, show different trendlines than the F, G, and K
stars. The M0 V -- M2.5~V stars show a shallower slope 
($\alpha = 1.76\pm 0.24$). The two early-M dwarfs with X-ray and
  Ly-$\alpha$ data obtained on the same day (GJ~644C and GJ~176)  are on the
  trendline established by stars with mostly non-contemporaneous data. 
The most active of the early-M stars, the young stars
AU~Mic (M0~V) and HIP~23309 (M0~V),
are at the top of the early-M dwarf and the F-G star trendlines. 
The least active stars, GJ~667C (M1.5~V) and 
GJ~176 (M2.5~V), have Lyman-$\alpha$ fluxes a factor of 10 times weaker than
the least active G and K stars with similar X-ray fluxes. 

The M3--M7.5 group also shows a shallow slope
($\alpha = 1.42\pm 0.17$) with the more active of these late-M stars having
Lyman-$\alpha$ fluxes a factor of 4 times lower
than the G and K stars with similar
X-ray fluxes. The least active of the M3--M7.5 dwarf stars 
(GJ~581 and GJ~436) have a factor of 10 times weaker Lyman-$\alpha$ flux 
than the G and K stars with similar X-ray fluxes. The coolest M
dwarf in our list, TRAPPIST-1 (M7.5~V), has a factor of 150 times lower
Lyman-$\alpha$ flux than the G and K stars with similar X-ray fluxes.
Figure~2 shows this pattern of decreasing chromospheric emission compared to
coronal emission for the increasingly cool and less active 
M dwarf stars.

The most active stars lie near the top of each
trendline and the least active stars lie near the bottom. 
For example, the two young stars near the top of
the G star trendline, HR~6748 (age $440\pm190$ Myr) and V993~Tau (age
$630\pm50$ Myr), have rotation periods of
5.9 and 4.65 days, respectively, whereas the old stars with the
lowest X-ray fluxes
at the bottom of the G star trendline, $\alpha$~Cen~A (age $5.3\pm0.3$
Gyr) and $\tau$~Cet (age $5.6\pm1,2$ Gyr) have
rotation periods of 28 and 34.5 days, respectively. Note that the
quiet Sun (G2~V)
lies near the bottom of the G star trendline close to $\alpha$~Cen~A
and $\tau$~Cet.
A star near the top of the K star trendline is Speedy Mic, a young 
($30\pm 10$ Myr) rapidly rotating star with a rotational period of 0.38 days. At
the bottom end of the K star trendline is the old ($6.9\pm0.4$ Gyr) star
HD~40307 with a rotation period of 48 days. For the M0~V -- M2.5~V
star group, AU~Mic (age $20\pm 10$ Myr) lies at the top of the trendline
and Kapteyn's star (age $11.5^{+0.5}_{-1.5}$ Gyr) lies at the bottom of
the trendline. A similar trend is seen for the M3~V -- M7.5~V star group
with the active and probably young 
stars AD~Leo and EV~Lac at the top of
the trendline and the $4.2\pm 0.3$~Gyr star Ross~905 (GJ~436) and the
$7.6\pm 2.2$~Gyr star TRAPPIST-1 near the
bottom of the trendline. The stellar ages are from the compilation of
\cite{Schneider2019}.

As stars age on the main
sequence, rotate more slowly, and generate less magnetic flux, they
descend the trendline for their spectral type. However, the time scale
for decreasing rotation depends on stellar mass with the F--K dwarfs
becoming slow rotators before the age of the Pleiades (125 Myr),
and the M3--M7.5 stars remaining rapid rotators at the age of Praesepe
(790 Myr) \citep{Rebull2017}.
Since X-ray fluxes and Lyman-$\alpha$ fluxes are usually not
measured at the same time for a given star, part of the scatter about
the trendlines is due to variable activity, which is larger for the M
stars than for the warmer stars \citep{Marino2002,Miles2017,Loyd2014}.

Figure 3 shows the decline of M dwarf chromospheres in a different
way. The figure plots reconstructed Lyman-$\alpha$
flux divided by the Lyman-$\alpha$ flux ratio predicted if the M
stars followed the trendline for the G stars at the same  X-ray
flux. The most important trend in this figure is the large decrease
in Lyman-$\alpha$ flux toward the later spectral types.

\subsection{Error Analysis for the Regression Coefficients}

We developed two methods for estimating the linear regression coefficients
and their uncertainties for the plots of log $F_x$ vs log $F_{Ly-\alpha}$
(Figure 2) and log $L_x/L_{\rm bol}$ vs log $L_{\rm Ly-\alpha}/L_{\rm bol}$.
The first method estimates the error of the regression fit by
bootstraping the
residuals. For the observed data $(x_i,y_i)$, we first estimate the
regression coefficients of the linear fit and calculate the fitted values
$\hat{y}$. The residuals of the fit are defined as $e_i=y_i-\hat{y}$. We then
built 10000 boostrap samples of the residuals $\widetilde{e_i}$. For each
residual series, we computed the boostrap
$\widetilde{y_i}=\hat{y}+\widetilde{e_i}$ and then obtained the linear
fit for each dataset $(x_i,\widetilde{y_i})$. In this way, we obtained the
mean and the standard deviation of the regression coefficients. These
results are shown in Table~2.

The second method, which is not a traditional bootstrap, involves allowing
both $x_i$ and $y_i$ to vary within a wide range and then solving for the
regression
coefficients multiple times to determine their mean values and uncertainties.
We allowed all values of $F_{\rm Ly-\alpha}$ or
$L_{\rm Ly\alpha}/L_{\rm bol}$ to vary randomly between 0.7 and 1.3 times the
observed values and all values of $F_x$ or
$L_x/L_{\rm bol}$ to vary randomly between zero and twice
the observed values. The results are very similar to those obtained by
the first method.

\section{Trend Lines for Luminosity divided by Bolometric Luminosity}

Between the early-F and late-M dwarfs, stellar effective temperatures,
radii, and bolometric luminosities change by large factors.
Ratios of X-ray and Lyman-$\alpha$ luminosities to the stellar
bolometric luminosities measure the relative amounts
of radiated energy from the chromosphere and corona. Table~1
lists bolometric luminosities computed from the stellar radii,
effective temperatures and distances listed in \cite{Schneider2019},
\cite{Melbourne2020}, and Youngblood et al. (in prep.). The ratios
R(Ly-$\alpha$)=$10^5$L(Ly-$\alpha$)/L(bol) and
R(X)=$10^5$L(X)/L(bol) in Table~1 are computed from the fluxes
and bolometric luminosities. The ages listed after the star names are in three
catagories: young stars (Y) 0--450~Myr, middle age stars (M) 0.5--3~Gyr,
and old stars (O) $>4$~Gyr. We group the stars into these three groups
as many stellar ages are imprecise.
Most of the ages are from the compilations of
\cite{Schneider2019} and \cite{Melbourne2020}.

In Figure~4 we plot  L(X)/L(bol) vs L(Ly-$\alpha$)/L(bol) for
the stars grouped by spectral type. Unlike Figure~2, the
stars in Figure~4 are clumped together as the small bolometric
luminosities of the M stars move their luminosity ratios to the
upper right in Figure~4. The regression line slopes $\alpha$  
in the luminosity ratio equation 
log L(X)/L(bol) = $\alpha$ log(L(Ly$\alpha$)/L(bol)) + $\beta$
are similar to the regression line slopes for the flux-flux equation as
shown in Table~2. GJ~176 is on the trendline for early-M stars, but
  GJ~667C. The errors in the regression coefficients were computed as
  described in Section 3.1.
We next consider how L(X)/L(bol) and L(Ly-$\alpha$)/L(bol) change with
stellar effective temperture ($T_{\rm eff}$) for stars with different ages..

\subsection{Young stars}

We have identified 21 young stars with
reconstructed Ly-$\alpha$ and X-ray fluxes.
The range in ages for these young stars is
0.02 Gyr (AU Mic and HIP~23309) to 0.44 Gyr (HR 6748).
The boundary between young and
middle aged stars is arbitrary, but separating young from middle-aged stars
at 0.45 Gyr reveals a clear pattern. We include two spectroscopic binaries
(V471~Tau and GJ 644B) in the young star list as tidally induced rapid
rotation raises their
activity levels to those of the rapidly rotating
young stars \citep{Wheatley1998}.
The distribution of data points in Figure~5 and least-squares fits to the
data show two interesting trends. 
The L(Ly-$\alpha$)/L(bol) data increase smoothly from
$T_{\rm eff} = 6,000$~K (near spectral type G0~V) to 3,000~K
(near spectral type M5~V), implying that the fraction of the stellar
bolometric luminosity that heats chromospheres increases steadily to cooler
stars. 

The L(X)/L(bol) data behave differently than L(Ly-$\alpha$)/L(bol).
At high temperatures, L(X)/L(bol) is roughly equal to L(Ly-$\alpha$)/L(bol),
but beginning near 5,400~K (about spectral type G8~V), L(X)/L(bol)
rises steeply to near the saturation value of $10^{-3.1}$.
L(Ly-$\alpha$)/L(bol) does not reach saturation for
young stars until $T_{\rm eff}\approx 3,000$. This behavior may result from the
different ages of stars in our young star sample. There are 12 stars in
Figure~5 with $T_{\rm eff}$ in the range 5167--5932~K. Eleven of these stars
have ages 200--450~Myr with a mean age of 310~Myr. All of these stars have
L(X)/L(bol) below $10^{-4}$. The exception is LQ Hya with age 70~Myr and
$T_{\rm eff} = 5376$~K, which has L(X)/L(bol) well above $10^{-4}$. Three other
somewhat cooler stars with ages 30--150~Myr have L(X)/L(bol) close to
$10^{-3.1}$ as shown in Figure~5. However, the L(Lyman-$\alpha$)/L(bol) for
these four stars are close to the least-squares fit. 
We interpret these data in terms of coronal heating in G and K stars
younger than about 150 Myr and cooler than about 5,200~K being enhanced
relative to chromospheric heating.

\subsection{Middle age stars}

For the 17 stars with middle ages, which we consider here to be 0.5~Gyr
($\epsilon$~Eri) to 3~Gyr ($\pi$~Men), the dependence on $T_{\rm eff}$ is
somewhat different than the young stars. Like the young stars,
L(Ly-$\alpha$)/L(bol) in Figure~6 also increases smoothly to lower
effective tempertures.
On the other hand, the L(X)/L(bol) data points are widely scattered with no
apparent pattern except that they are far below saturation and well below
L(Ly-$\alpha$)/L(bol).

\subsection{Older stars}

In Figure~7 we show data for the group of 33 older stars between ages
4.0~Gyr (GJ 176) and
11.6~Gyr (Kapteyn's star). The data for GJ~176 are on both trendlines.
Similiar to the young and middle age stars,
L(Ly-$\alpha$)/L(bol) also increases smoothly to the coolest stars in our
sample. Although the L(X)/L(bol) data are scattered, they show
an increasing trend to lower $T_{\rm eff}$ with a steeper slope than
L(Ly-$\alpha$)/L(bol). The mean ratio of L(Ly-$\alpha$)/L(bol) to
L(X)/L(bol) decreases from a factor of about  100 near $T_{\rm eff}=6,000$~K
to a factor of about 3 near
3,000~K. L(X)/L(bol) and L(Ly-$\alpha$)/L(bol) are nearly equal for some
of the coolest stars. 
We find that for the older stars, with decreasing $T_{\rm eff}$,
coronal emission and heating become increasingly important
  relative to chromospheric emission and heating.

\subsection{Comparison of L(Ly-$\alpha$)/L(bol) for stars of different age
  groups}

Figure 8 compares the the least squares fits to the L(Ly-$\alpha$)/L(bol) data
for the young, middle age, and older stars. These three plots are nearly
parallel and
show a decrease by a order of magnitude at all effective temperatures
between the young and older stars.
The black line in the figure is the least-squares plot of L(X)/L(bol) for
the older stars showing an increased slope toward lower $T_{\rm eff}$
compared to L(Ly-$\alpha$)/L(bol).

\subsection{Comparison of L(Ly-$\alpha$/L(bol) with L(X)/L(bol) for stars
  with saturated X-ray emission}

Young F, G, and K (FGK) stars are rapid rotators that
generally show maximum X-ray emission with
L(X)/L(bol) $\approx 10^{-3.1}$ \citep{Pizzolato2003}.
For early F stars, however,  the maximum value for L(X)/L(bol) is about
$10^{-4.3}$ \citep{Jackson2012}. The decrease
from saturated X-ray emission occurs with decreasing rotation rate
after about age 100 Myr for FGK stars but significantly later for M stars that
have longer spin-down times \citep{Newton2017}. For FGK stars,
saturated emission occurs when the rotation period is less than 1.3 to 3.5 days,
depending on stellar mass, but the maximum rotation period for saturation
is greater than about 10.8 days for M stars \citep{Pizzolato2003}.
As measured by the GALEX NUV and FUV fluxes corrected for
photospheric emission, chromospheric emission is also saturated until
approximently 100 Myr for FGK stars but much later ages for late-M stars
\citep{Schneider2018,Richey-Yowell2019}. GJ~871.1 (M4~IV), a young star with
L(X)/L(bol) exceeding $10^{-2}$, 
was not included in our analysis as the observed L(X)/L(bol)
ratio exceeded the saturation
level by more than an order of magnitude indicating that flares occured
during the observations.

Our data set provides an opportunity to explore the saturation behavior of
X-ray and Lyman-$\alpha$ emission as a function of $T_{\rm eff}$
in the same stars. We first select stars with ages
less than 100~Myr. As shown in Figure~9, there are 8 stars
that meet this criterion. For stars cooler than 5075~K (V368~Cep),
the values of L(X)/L(bol) are close to $10^{-3.1}$ with little dispersion,
but the corresponding
L(Ly-$\alpha$)/L(bol) values are much lower than L(X)/L(bol).
This shows that when the coronal emission is saturated
(or nearly so), the saturated emission level for chromospheric emission is
an order of magnitude lower for early K stars ($T_{\rm eff}\approx 5000~K$)
but only a factor of 3 lower for late-M dwarfs. 

To check on the broader applicability of these trends, we include older
stars with large L(X)/L(bol). These are shown as colored symbols in Figure~9.
PW And is a K2~V star in the AB Dor moving group with an age of
$149^{+51}_{-19}$~Myr. The values of both
L(X)/L(bol) and L(Ly-$\alpha$)/L(bol) are consistent with the younger stars
at similar $T_{\rm eff}$. The M4~V star YZ CMi also fits these trends
despite its field star age (about 5~Gy), indicating that late-M stars
can also have
saturated emission like the young stars. Another group of stars that could have
saturated X-ray and Ly-$\alpha$ emission are short period spectroscopic
binaries that have been spun-up due to tidal interactions. There are two
spectroscopic binaries in Table~1: the $P_{\rm orb}=0.521$ day period
V471~Tau (K2 V + DA) binary \citep{Hussain2006} and the
$P_{\rm orb}=2.9655$ day  period GJ 644B (M3.5V + SB) multiple system
\citep{Mazeh2001}.
The two systems are only partially consistent with the trends shown
by the younger stars. For V471~Tau, L(X)/L(bol) is slightly above $10^{-3.1}$ 
perhaps due to X-ray emission from the hot white dwarf star,
but L(Ly-$\alpha$)/L(bol) is far above the trend line. For GJ~644B.
L(X)/L(bol) is well below $10^{-3.1}$, but L(Ly-$\alpha$)/L(bol) is on the
trend line. A much larger sample of spectroscopic binaries is needed to
test whether their saturation properties are similar to the $<100$~Myr stars.

\section{Discussion}

\subsection{Trends with spectral type and effective temperature}

Compared to the F, G, and K stars, M dwarfs show an increasing trend of
weak chromospheric emission relative to their coronal emission, as clearly
shown in Figures~2 and 3. Stellar age also
plays an important role in determining whether
chromospheric emission becomes relatively weak or
coronal emission becomes relatively strong with decreasing $T_{\rm eff}$.
Since emissions from both chromospheres and coronae
decay with decreasing rotation rate and magnetic flux that occur as
stars age on the main sequence, we have separated stars into three age
groups: young (age $<450$~Myr), middle aged (0.5--3~Gyr) and older ($>4$~Gyr).
This crude age discriminant is needed
to have at least 15 stars in each group for decent statistics. 

The smooth increase of L(Ly-$\alpha$)/L(bol) with decreasing 
$T_{\rm eff}$ for stars in all age groups says that with decreasing
$T_{\rm eff}$ an increasing fraction 
of the stellar luminosity heats chromospheres irrespective of stellar age.
The decrease in chromospheric heating by a factor of 10 from the young to
the older stars shown in
Figure~8 is remarkable in that it shows that stellar effective temperature
and thus bolometric luminosity are relatively unimportant parameters for
understanding the relative decline of chromospheric emission with stellar age.
\cite{France2018} also found that the decrease in the normalized
flux for the
chromospheric Si~III 120.6~nm line, L(Si~III)/L(bol), with
rotation period is the same for F, G, K, and M stars. These two results are
consistent as the increase in rotation period is correlated with stellar age.

The trend of L(X)/L(bol) with $T_{\rm eff}$ and stellar age is more complex.
For the young star group, the increase in L(X)/L(bol) with decreasing effective
temperature overlaps L(Ly-$\alpha$)/L(bol) for F and G stars
($T_{\rm eff}= 5400-6300$~K), but L(X)/L(bol) then rapidly rises to
the saturation level
($10^{-3}$) while L(Ly-$\alpha$)/L(bol) continues its steady increase to
lower effective temperatures. This suggests that for G and K stars younger
than about 150~Myr, coronal heating exceeds chromospheric heating by about
an order of magnitude and this difference decreases with age. 
New observations are needed to test this interpretation of the relative
heating of coronae and chromospheres. 

For older stars there
is a simple trend, L(X)/L(bol) increases far more rapidly with decreasing
$T_{\rm eff}$ than does L(Ly-$\alpha$)/L(bol). Near $T_{\rm eff}$=6000~K,
L(Ly-$\alpha$)/L(bol) is 100 times larger than L(X)/L(bol), but
L(Ly-$\alpha$)/L(bol) is only a factor of 3 times larger than L(X)/L(bol) 
near $T_{\rm eff}=2500$~K. Thus for older stars there is a more  rapid
increase in coronal heating compared to chromospheric heating with
decreasing $T_{\rm eff}$.

The dispersion of L(X)/L(bol) about the trend lines is much
larger than the dispersion of L(Ly-$\alpha$)/L(bol) for the middle age and
older stars. In part this must be a consequence of the large variation in
X-ray flux between magnetic cycle maximum and minimum. For $\alpha$~Cen A and
B, \cite{Ayres2014} found minimum-to-maximum X-ray flux ratios of 3.4 and 4.5,
respectively, as observed
by the {\em Chandra}/HRC instrument with its 0.5--17.5~nm bandpass. These
minimum-to-maximum contrasts are larger that what is typically seen in
chromospheric lines. With its spectral bandpass limited to $<3.5$~nm, the
instruments on {\em XMM-Newton} detected a very deep minimum in the X-ray flux
from $\alpha$~Cen~A in 2005. This points to the second reason for the large
dispersion in the L(X)/L(bol) data for middle age and older stars.
Such stars have
coronae with temperatures near $1\times 10^6$~K \citep{Ayres2014}. Decreases
in coronal temperature near cycle minimum cause a larger fraction of the
coronal emission to be at the longer wavelengths that {\em XMM-Newton} and
even {\em Chandra} cannot detect. The combination of variations in coronal
heating over a cycle, changing spectral hardness compared to
instrumental bandwidth, and flares  conspire to produce the large scatter
in L(X)/L(bol) about the trendlines.

\subsection{Possible explanations for the different emissions from coronae and
  chromospheres}

We find that the relative emission and therefore
heating rates of chromospheres and coronae depend on stellar age and
effective temperature very differently. What could be the contributing factors
to this different behavior? We consider 
four important differences between M dwarfs and warmer stars that
could explain the different trendline slopes and the different
L(Lyman-$\alpha$)/L(bol) and L(X)/L(bol) trends with $T_{\rm eff}$:
(1) M stars have higher gravity and lower photospheric opacity,
(2) the different effects of flares on the Ly-$\alpha$ and X-ray emissions,
(3) M stars have increased molecular formation and
lower ionization in their photospheres and lower chromospheres, and
(4) the absence of a radiative 
core structure in stars cooler than spectral type M3.5~V. We now consider each
of these factors.

(1) Since $R_{\\star}^2$ decreases faster than stellar mass along the
main sequence, M dwarfs have higher gravities ($g\propto M/R_{\star}^2$)
than solar-mass stars by a factor of 2--3. 
Since the pressure scale height is inversely 
proportional to gravity, M dwarf photospheres are more compact than for
more massive stars. M dwarf photospheres are also denser than for
warmer dwarf stars, because important
opacities (e.g., H and H$^-$) are smaller due to lower
temperatures and decreased ionization. It appears that there may be a
fundamental link between atmospheric structure and the trendlines relating
heating of coronae and chromospheres. The precise nature of this link is
uncertain, but one component could be that higher
photospheric gas pressures increase the equipartition magnetic field
strengths ($P_{\rm gas}=B^2/8\pi$), and M dwarfs usually have significantly 
higher magnetic field strengths than solar type stars
\citep[e.g.,][]{Reiners2012}. We consider this to be the most interesting of
the possible explanations.

(2) For three older M dwarfs, \cite{Loyd2018} found
that flares contribute at least 10--40\% of the
quiescent FUV flux, and that during
flares the Ly-$\alpha$ flux increases by a much smaller factor than
higher temperature lines and the X-ray flux. Since flares are more
frequent in older M dwarfs than older G dwarfs, flaring may
play a role in
explaining the relative behavior of X-ray and Ly-$\alpha$ emission with
decreasing $T_{\rm eff}$. Coronal temperatures determine the fraction of the
input magnetic energy that is radiated within the 
{\em XMM-Newton} and {\em Chandra} bandpasses as opposed to the
extreme-UV wavelengths outside of the X-ray bandpasses. The cooler coronae of
older G stars like the Sun have a higher percentage of coronal emission in the
EUV range than the hotter coronae of older M dwarfs.
The more rapid increase in L(X)/L(bol) compared to
L(Ly-$\alpha$)/L(bol) with decreasing $T_{\rm eff}$ could be due in part to
the higher percentage of coronal emission in the {\em Chandra} and
{\em XMM-Newton} bandpasses rather than in the EUV with decreasing
$T_{\rm eff}$. A quantitative assessment of this effect would be useful but is
beyond the scope of this paper.

(3) M stars have lower temperatures in their photospheres and lower
chromospheres compared to warmer stars, as shown for example by the
temperature distributions of GJ~832 (M2~V) and the Sun in Figure~1 of
\cite{Fontenla2016}. Lower temperatures lock up atoms in molecules and 
cause reduced ionization.
With decreasing photospheric temperatures, hydrogen and abundant
metal atoms (e.g., C, N, and O) can be sequestered in diatomic and more
complicated molecules. At these low temperatures only atoms with
very low ionization potentials such as Na and K can supply free electrons
needed for important opacity sources such as H$^-$,
but these atoms have very low abundances. Also, the gases in
nearly neutral photospheres and lower chromospheres are poor
electrical conductors, which may
alter MHD wave dissipation and flare heating processes in
chromospheres. The heating in stellar coronae, however, may be unaffected as
hot plasma is highly ionized.

(4) A fourth possible cause for relatively weak chromospheric emission
compared to coronal emission may be related 
to the switch from radiative cores in stars warmer than spectral type M3.5~V 
to fully convective interior structures of cooler stars.
Interior structure models \citep[e.g.,][]{vanSaders2012,Baraffe2018}
predict that after stars reach the main 
sequence, stars with masses greater than about $0.35M_{\odot}$
acquire and retain 
radiative cores, whereas stars with lower masses remain fully convective.
At $M_{\star}=0.35M_{\odot}$ dwarf stars have
$T_{\rm eff}\approx3,300$~K and spectral type near M3.5~V.
\cite{Baraffe2018} found that near this critical mass, the abundances of $^3$He
in the stellar core and envelope play important roles in nuclear reaction
rates leading to a subtle change in stellar luminosity and color seen 
in {\em GAIA} photometry \citep{Jao2018}.

For solar-type stars, the regeneration and amplification of magnetic
fields is generally described by a $\alpha\Omega$-type dynamo
\citep{Dobler2005,Cameron2017}. In the
simplest example of a kinematic $\alpha\omega$ dynamo, the velocity fields are
specified rather than self-consistently computed.
The $\alpha$ effect converts torroidal
magnetic fields into poloidal fields, and the $\Omega$ effect converts
poloidal fields into torroidal fields. The interplay
between these two processes is driven by stellar rotation and turbulence. 
The regeneration of magnetic fields in $\alpha\omega$ dynamo models 
is usually thought to occur mainly near the tachocline, the interface
between the radiative core and convective envelope,
where rotational shear and associated turbulence are
a maximum. Stellar dynamos
are likely far more complex than this simple model, and many
problems are still inadeqately understood \citep{Charbonneau2014,Brun2017}

Without a radiative core, a star has no tachocline and the simple
kinematic $\alpha\omega$ type dynamo should no longer be feasible.
However, fully convective cool M dwarfs do have strong magnetic fields
\citep{Reiners2012}  indicating that magnetic heating is occuring in 
their chromospheres and coronae. A different type of
dynamo, perhaps a $\alpha^2$-type dynamo, could operate in the convective
envelopes of stars cooler than M3.5~V  \citep{Chabrier2000} and
perhaps also in the warmer M stars. 

\subsection{Search for observational evidence of changes when stars
  become fully convective}

While there have been  many searches for changes in the X-ray, UV, 
and H$\alpha$ fluxes that
could result from a fundamental change in the way magnetic fields
are amplified in stellar interiors, most have found no changes near
spectral type M3.5~V.
For example, \cite{Stelzer2013} did not find
any obvious changes in the ratios of fluxes in different energy bands
between early and late-M
dwarfs that would indicate a change in interior structure. In their 
volume-limited survey of X-ray emission from K and M dwarf stars,
\cite{Fleming1995} also found no evidence for a decrease in L(X)/L(bol)
and thus coronal heating efficiency between early-M dwarfs and the
fully convective late-M dwarfs. 
\cite{Loyd2018} found that the rate of flaring is the same for both
young and old M dwarfs with no obvious spectral type dependence
within the M dwarfs.

In their study of M dwarfs with masses
0.1--0.5 $M_{\odot}$, \cite{Newton2017} found that there is a single
relationship between L(H$\alpha$)/L(bol) and the Rossby number
$R_0=P_{\rm rot}/\tau$, where $P_{\rm rot}$ is the stellar rotation
period and $\tau$ is the convective turnover time,
implying no change in the magnetic dynamo across the
radiative/convective core boundary. \cite{Wright2016} showed that
fully convective M dwarfs show the same behavior of L(X)/L(bol)
as seen in the more massive stars
and concluded that both the more massive M
stars with radiative cores and the less massive fully convective stars
likely have the same dynamo process, perhaps a dynamo that generates
magnetic fields by helical turbulence first proposed by \cite{Durney1993}.

Are there any observations of different magnetic field properties that
could identify changes in stellar interior structures?
\cite{Shulyak2015} found no difference in the magnetic
field distributions between partially and fully convective stars in their
high resolution study of Zeeman broadening of spectral lines, and
\cite{Morin2010} found that the fully convective M dwarfs have a diverse
range of magnetic topologies with a tendancy for a higher degree of
magnetic field organization (more flux in lower degree sperical harmonics)
than the partially convective stars.

Magnetic fields in stellar coronae likely
play a major role in the first ionization potential (FIP) effect
in which elements with first ionization potential (FIP) less than 
about 10~eV are enhanced relative to elements with FIP greater than hydrogen
(13.5~eV). First identified in the solar corona
\citep[e.g.,][]{vonSteiger1995,Feldman2000,Laming2015},
the FIP effect has also been
identified in stellar coronae \citep[e.g.,][]{Drake1997,Gudel2001}.
In their analysis of {\em Chandra} Low Energy Transmission
Grating Spectrograph (LETGS) data, \cite{Wood2018} showed that the FIP effect
occurs in A7~V to K5~V stars, and for
stars cooler than K5~V including M dwarfs, an inverse-FIP effect
occurs when the low-FIP elements have lower coronal
abundances than the high-FIP elements. This change from FIP to inverse-FIP
may indicate that the propagation of MHD
waves through coronal magnetic loops is very different in the M stars
compared to warmer stars \citep{Wood2018}. However, the change
occurs near spectral type K5 ($T_{\rm eff} \approx 4400$~K) not M3.5
($T_{\rm eff} \approx 3300$~K).

While we find for the older stars both L(X)/L(bol)
and L(Lyman-$\alpha$)/L(bol) increase smoothly (with different slopes)
to lower effective temperatures, we find no clear evidence for changes
in either quantity near the fully convective bouldary. A much larger data set is
needed to find any substantial changes near this boundary.

\section{Conclusions}

This paper seeks to answer two questions that are not usually asked in
previous studies of dwarf stars with convective interiors. 
First, what is the relative amount of UV flux from
stellar chromospheres and X-ray flux from
stellar coronae emitted by F, G. K. and M dwarfs stars?
These emissions as measured by the stellar Lyman-$\alpha$
flux and the X-ray flux, respectively, are critically important
for understanding the rate of mass
loss and photochemical reactions occuring in exoplanet atmospheres. 
Are there patterns in the relative behavior of
these two types of emission for different types of stars? 

Second, we ask whether the relative amount of heating in
stellar chromospheres and coronae depends upon stellar effective 
temperature and age? The answer to this question is important for
determining which types
of heating processes operate in stars. 
Our analysis of 79 stars observed in recent observing
programs with {\em HST} and several X-ray
observatories has led to the following conclusions:

\begin{itemize}
\item For stars with similar
  spectral types and effective temperatures, the trendlines
  between chromospheric and coronal emission are described by power laws.
  As stars become less active with increasing age,
  slower rotation, and weaker magnetic fields, the relative
  fluxes of the chromospheric and coronal emissions both decrease following
  the trendline for their spectral type. 
  We show that the trendlines
  for F, G, and K dwarfs are nearly identical, but the trendlines
  for M stars differ significantly from those of the warmer
  stars. In particular, the trendline slopes ($\alpha$) for the
  M0--M2.5 stars ($\alpha=1.76\pm 0.24$) and for the M3--M7.5 stars
  ($\alpha=1.42\pm0.17$) 
  are smaller than for the F2--G9~V stars ($\alpha=2.32\pm 0.26$) and
  K0--K7~V stars ($\alpha=2.34\pm 0.31$).

\item At the same X-ray flux level (normalized to a distance of 1~au),
  the more active M3--M7.5 stars show Lyman-$\alpha$
  emission a factor of 4 smaller than corresponding F--K stars, and for
  the least active late-M stars the Lyman-$\alpha$ emission is
  a factor of 10 lower.
  The coolest star in the sample, TRAPPIST-1 (M7.5~V), emits a factor of
  150 times less Lyman-$\alpha$ flux than does the least active
  F--K stars with similar X-ray fluxes.
  The relative amounts of chromospheric and coronal emission
  in M stars are thus qualitatively different from the warmer stars. This
  difference is most extreme for the coolest star in the sample.

\item We call attention to a possible fundamental link between atmospheric
  structure and the trendlines relating heating of coronae and chromospheres.
  With their higher gravities and lower photospheric temperatures, M dwarf
  atmospheres are denser, less ionized, and have smaller pressure scale
  heights than warmer stars. The precise nature of the link is
  uncertain, but one underlying cause could be that higher
  photospheric gas pressures increase the equipartition magnetic field
  strengths ($P_{\rm gas}=B^2/8\pi$), and M dwarfs usually have significantly 
  higher magnetic field strengths than solar type stars. The relative
    heating rates in chromospheres and coronae should be related to the
    strength and scale height of the magnetic fields in stratified atmospheres
    with different pressure scale heights.
  
\item The M stars show different trendlines and wider scatter
  about the trendlines than the warmer stars. Increased scatter is
  expected as even low activity M dwarfs flare more often
  than warmer stars and show larger UV and
  X-ray variability. The different trendlines for the M stars
  compared to the warmer stars suggest that
  different heating mechanisms may operate in the M stars compared to the 
  warmer stars that distribute heat differently between their
  chromospheres and coronae.

\item We find that the fraction of a star's bolometric luminosity that heats
  its chromosphere increases smoothly
  with decreasing $T_{\rm eff}$ for dwarf stars of all ages, but this fraction
  decreases a factor of 10 from
  the group of young stars (age $<450$ Myr) to the older stars (age $>4$ Gyr)
  at all effective temperatures.

\item The dependence of L(X)/L(bol) on $T_{\rm eff}$ is very different for
  the young and older stars. For G and K stars younger than about 150~Myr,
  saturated coronal heating as measured by L(X)/L(bol) is significantly
  larger than saturated chromospheric heating as measured by
  L(Lyman-$\alpha$)/L(bol). As stars age, both L(X)/L(bol) and
  L(Ly-$\alpha$)/L(bol) decrease from their saturation levels, and
  L(Ly-$\alpha$)/L(bol) becomes larger than L(X)/L(bol). 
  For the older stars, L(X)/L(bol) increases far more steeply with
  decreasing $T_{\rm eff}$ than does L(Ly-$\alpha$)/L(bol). We conclude that
  at least for the older stars, coronal heating becomes much more important
  compared to chromospheric heating with decreasing $T_{\rm eff}$.
  
\item We asked whether the decreasing ratio of Lyman-$\alpha$ to
  X-ray emission in the cooler M dwarfs results from chromospheres
  becoming weaker or coronae becoming stronger. Plots of L(X)/L(bol)
  and L(Lyman-$\alpha$)/L(bol) with respect to stellar effective temperature
  show that for the younger stars 
  chromospheric emission increases gradually with
  decreasing effective temperature, but the coronal emission increases
  dramatically to saturation levels for stars with $T_{\rm eff}<5,300$~K
  (spectral type K2~V). The likely explanation for this effect is that for
  stars younger than about 150~Myr, coronal heating is an order of magnitude
  larger than chromospheric heating, but this difference decays with age.
  For the older stars, L(X)/L(bol) increases much faster to lower
  $T_{\rm eff}$ than does L(Lyman-$\alpha$)/L(bol). We have described several
  possible explanations for coronal heating increasing much faster than
  chromospheric heating with decreasing $T_{\rm eff}$. We
  conclude that the decreasing ratio of Lyman-$\alpha$ to
  X-ray emission in the cooler M dwarfs results from coronal emission becoming
  stronger rather than chromospheric emission  becoming weaker with decreasing
  $T_{\rm eff}$.

\item We searched for observational evidence for an abrupt change in
  Lyman-$\alpha$ or X-ray flux near spectral type M3.5~V
  ($T_{\rm eff}\approx 3,300$~K) where dwarf stars become fully convective.
  We found gradual changes in Lyman-$\alpha$ and X-ray fluxes and luminosities
  devided by bolometric luminosity but no abrupt changes at or near the
  boundary.
  
\item Photochemistry and mass loss in exoplanet atmospheres are driven
  by the spectral energy distribution of the host star's
  radiation. Very different evolutions of exoplanet atmospheres can
  result from whether the host star is an M dwarf or a warmer star.
  The different trendlines of M stars compared to warmer stars 
  mean that the spectral energy distributions of M stars will evolve in
  different ways than warmer stars as stars age on the main
  sequence. This difference in the host star's evolution
  will be important in assessing whether exoplanets of M stars can
  retain their atmospheres.  

\end{itemize}

\acknowledgments
JLL acknowledges support from STScI for programs HST-AR-15038, HST-GO-15071,
HST-GO-15190, and HST-GO-15326.
BW acknowledges support from the STScI and NASA through program
HST-GO-15326. AY thanks STScI and NASA for funding program HST-GO-15190.
A. Brown acknowledges support for processing X-ray observations used
  in this paper from {\em Chandra} Guest Observer grants GO4-15014X,
  GO5-16155X, and
GO8-19017X. KF acknowledges support through HST-GO-15071. SP thanks STScI
and NASA for support of program HST-GO-14640. SR thanks the Glasstone
Foundation and Jesus College for her research funding and support.
PW acknowledges support from STFC through consolidated grants
ST/P000495/1 and ST/T000406/1.

\facilities{HST(STIS), HST(COS), XMM-Newton, Chandra X-ray
  Observatory, ROSAT}

\clearpage
\begin{figure}[t]
\begin{center}
\includegraphics[angle=0,scale=1.0]{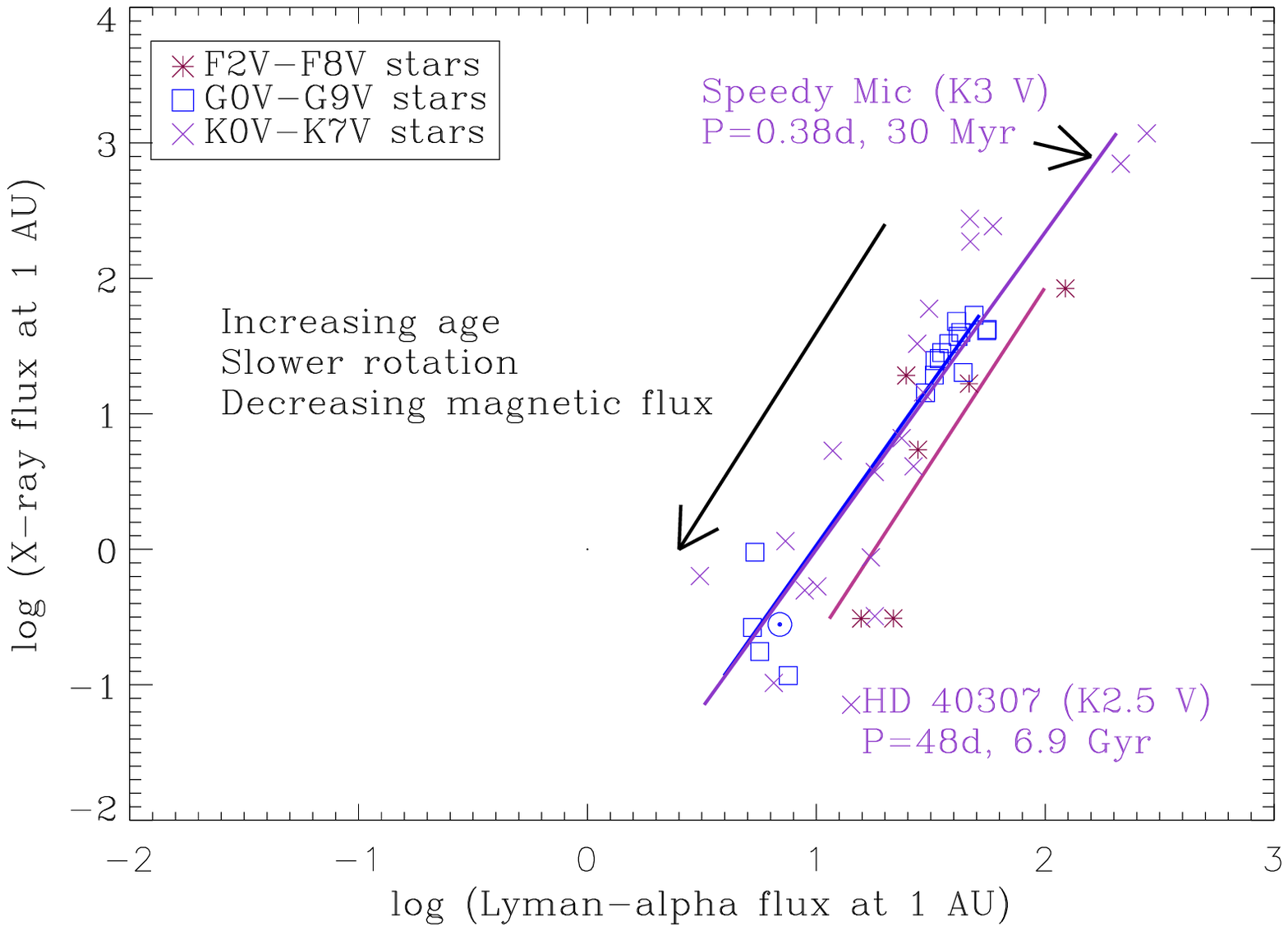}
\end{center}
\caption{A plot of the X-ray flux vs. the reconstructed Lyman-$\alpha$ flux
  for stars at a standard distance of 1 au. Different symbols
  represent the fluxes of F (maroon), G (blue), and K (violet) dwarf stars.
  The solid lines with the same colors are least square fits to the data
  for stars in each spectral type category. The names,
  rotational periods, and ages of the very active K star (Speedy Mic) and
  the least active K star (HD~40307) are identified. The
  $\odot$ symbol is the Sun at low activity.}  
\end{figure}

\clearpage
\begin{figure}[t]
\begin{center}
\includegraphics[angle=0,scale=1.0]{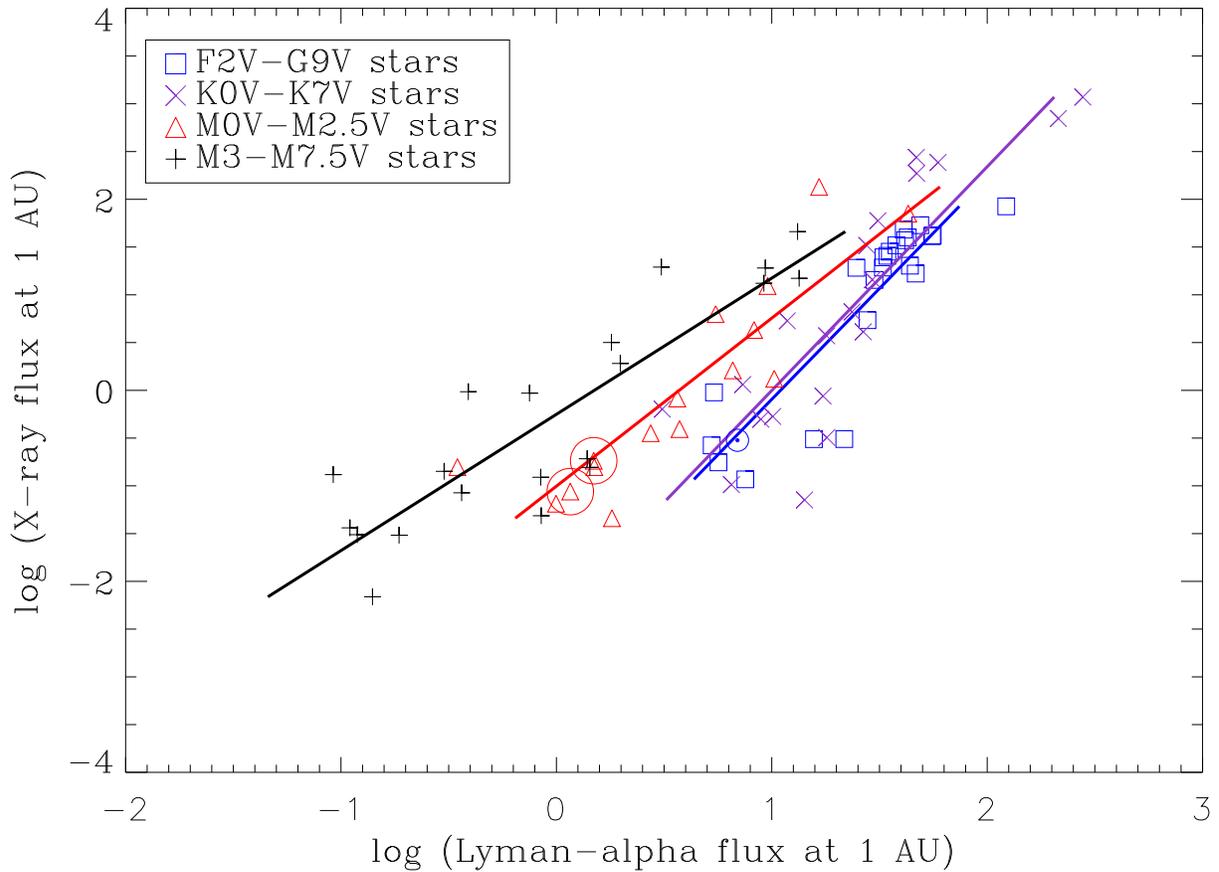}
\end{center}
\caption{A plot of the X-ray vs. reconstructed Lyman-$\alpha$ flux 
  at 1 au for dwarf stars. Different symbols
  represent the fluxes of F and G (blue), K (violet),
  M0--M2.5 (red), and M3--M7.5 (black) stars.
  The solid lines with the same colors are least square fits
  to the stars in each spectral type category. The 
  metal-poor subdwarf Kapteyn's star (M1~VI) is the early M star with the
  smallest Lyman-$\alpha$ flux and the X-ray brightest early-M star is
  the very young star HIP~23309 (M0~V). The two circuled symbols are for the
  stars with X-ray and UV observations obtained within less than one day:
  GJ~176 (above) and GJ~667C (below). The $\odot$ symbol is the Sun
  at low activity.}
\end{figure}

\clearpage
\begin{figure}[t]
\begin{center}
\includegraphics[angle=90,scale=0.75]{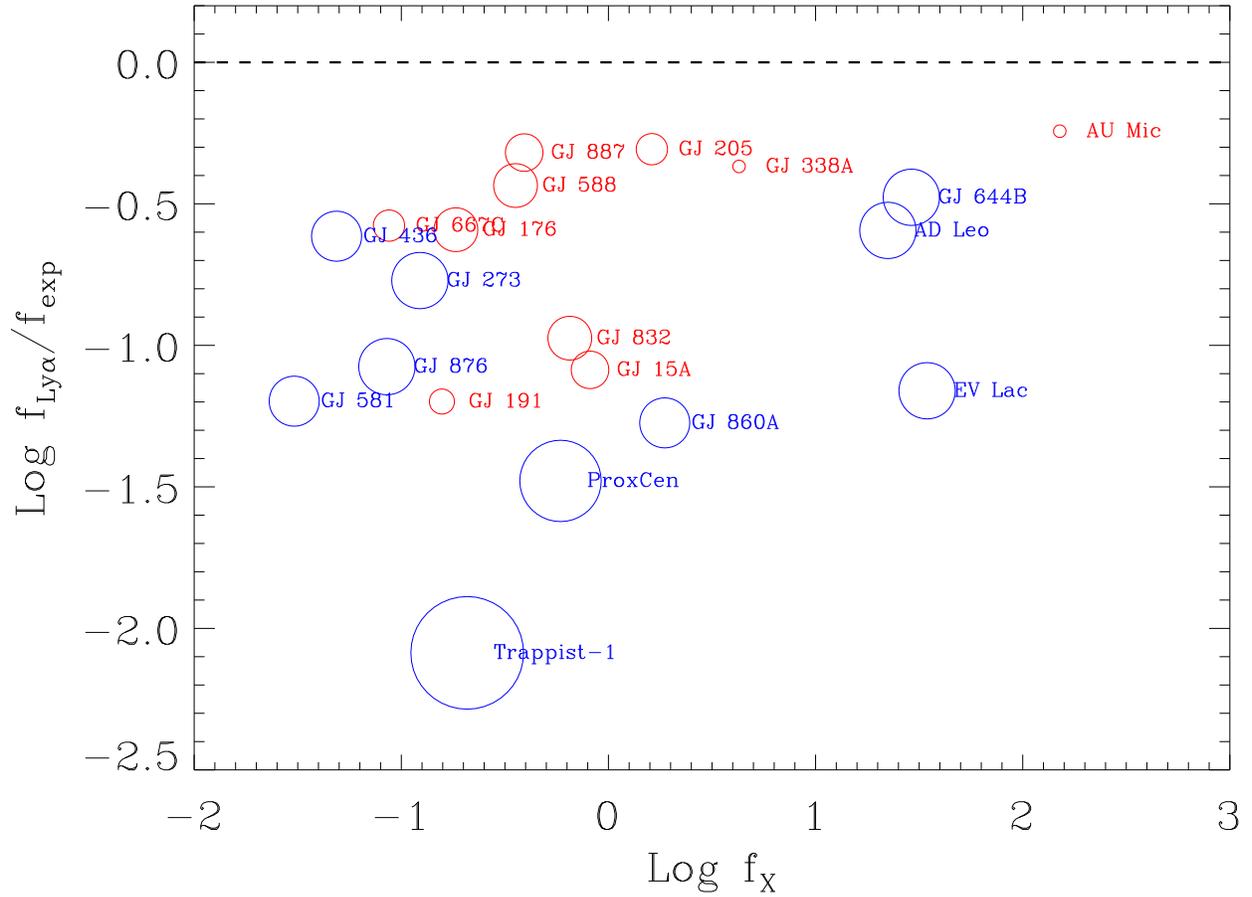}
\end{center}
\caption{Ratios of the reconstructed Lyman-$\alpha$ flux,
  $f_{\rm Ly\alpha}$ to the Lyman-$\alpha$ flux predicted if the M
  stars followed the G star trend line, $f_{\rm exp}$, at the same X-ray flux.
  The stars are indentified and color coded red for M0~V to M2.5~V
  stars or blue for M3~V to M7.5~V stars. The larger circles indicate
  later spectral type. The horizontal dashed line is
  where the M stars should lie if their Lyman-$\alpha$ flux followed
  the G star trend line for the same X-ray flux.}
\end{figure}

\clearpage
\begin{figure}[t]
\begin{center}
\includegraphics[angle=0,scale=1.0]{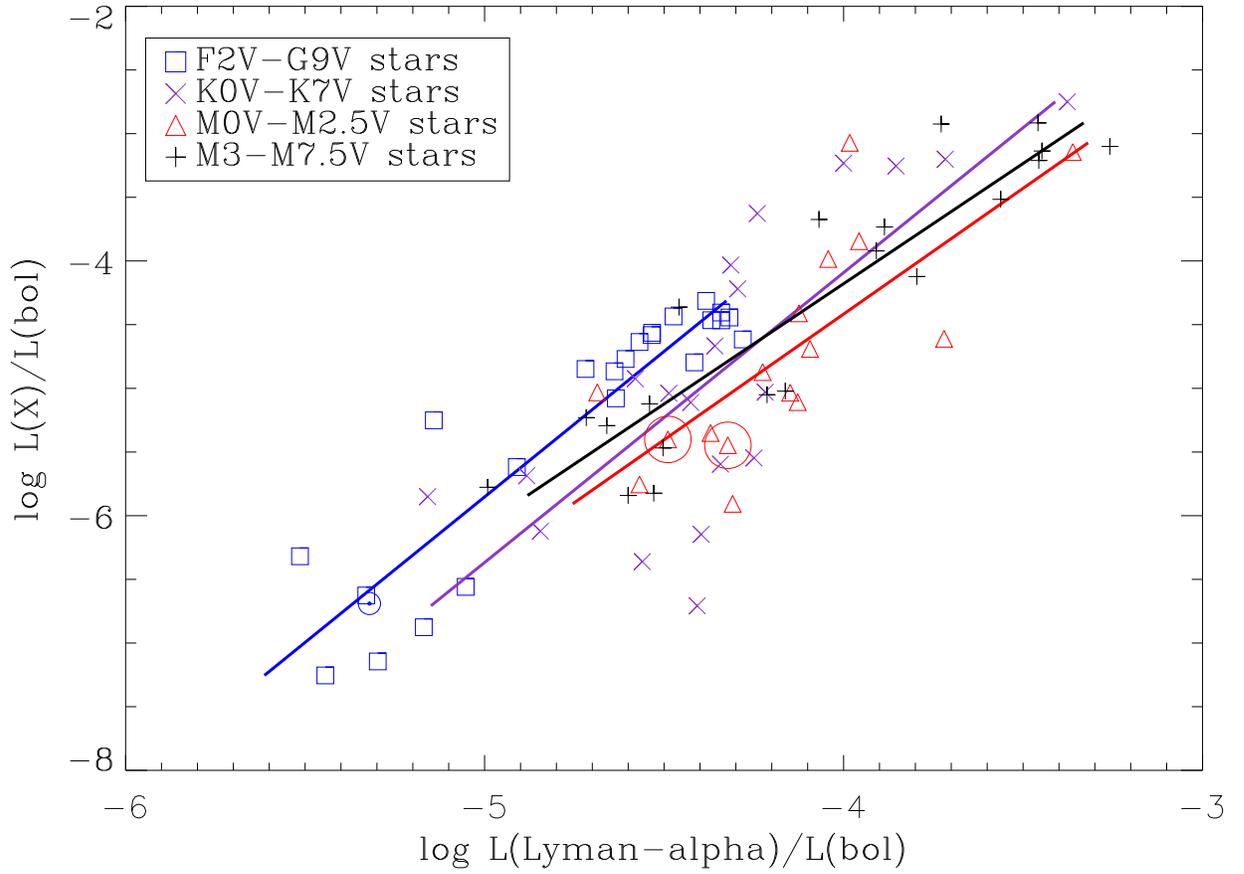}
\end{center}
\caption{A plot of the X-ray to bolometric luminosity ratios vs the
reconstructed Lyman-$\alpha$ to bolometric luminosity ratios. 
The data points are color coded with different symbols for the F2--G9 (blue),
K (violet), M0--M2.5 (red), and M3--M7.5 (black) stars.
The solid lines with the same color coding
are least-squares linear fits to the data in each spectral type range.
The two circuled symbols are for the
  stars with X-ray and UV observations obtained within less than one day:
  GJ~176 (left) and GJ~667C (right).
The quiet Sun is identified by the $\odot$ symbol.} 
\end{figure}

\clearpage
\begin{figure}[t]
\begin{center}
\includegraphics[angle=0,scale=1.0]{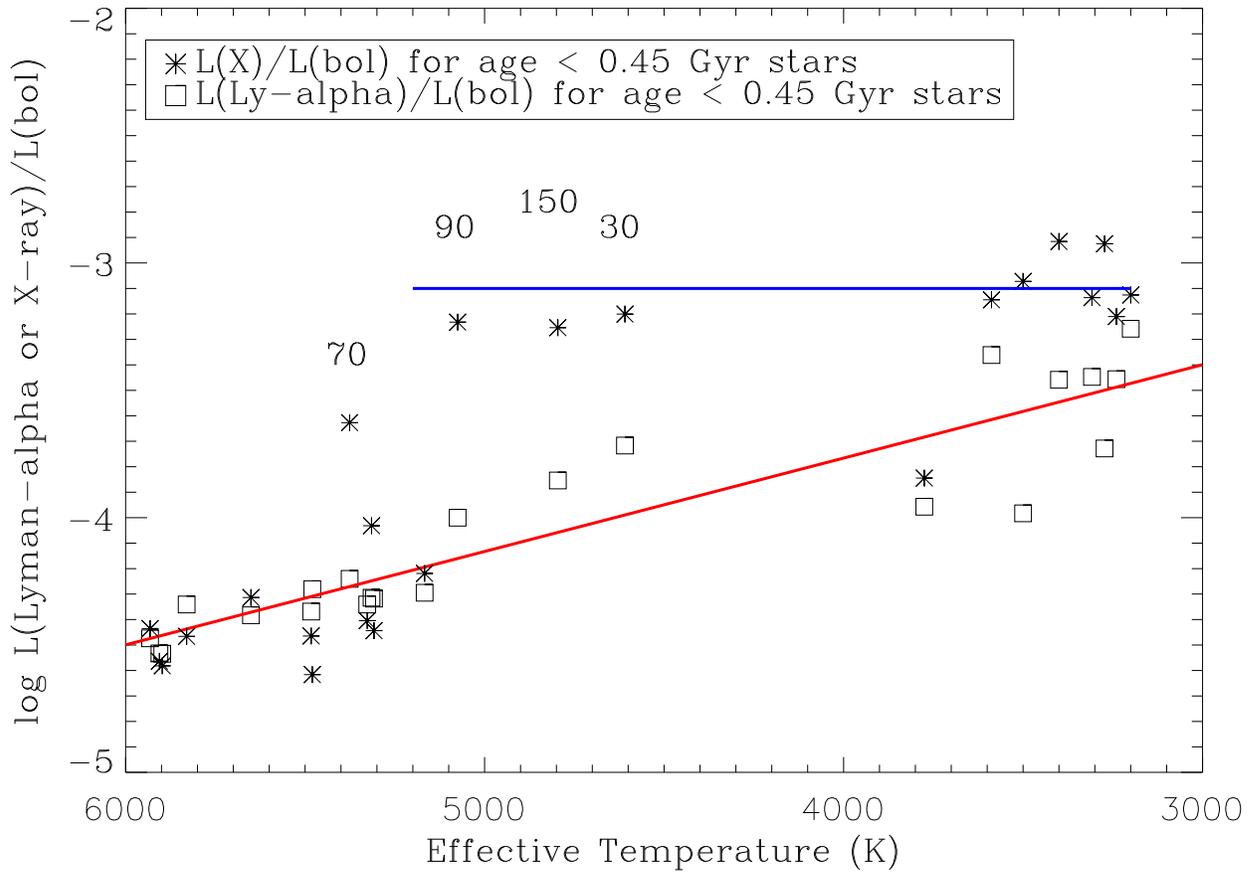}
\end{center}
\caption{Plot of L(X)/L(bol) and L(Ly-$\alpha$)/L(bol) vs. effective
  temperaturefor stars
  younger than 450 Myr. The solid red line is a least-squares linear
  fit to the L(Ly-$\alpha$)/L(bol) data. The solid blue line is the $10^{-3.1}$
  saturation level. The ages (in Myr) of four stars with temperatures
  between 4600 and 5400~K are given above the stars.}
\end{figure}

\clearpage
\begin{figure}[t]
\begin{center}
\includegraphics[angle=0,scale=1.0]{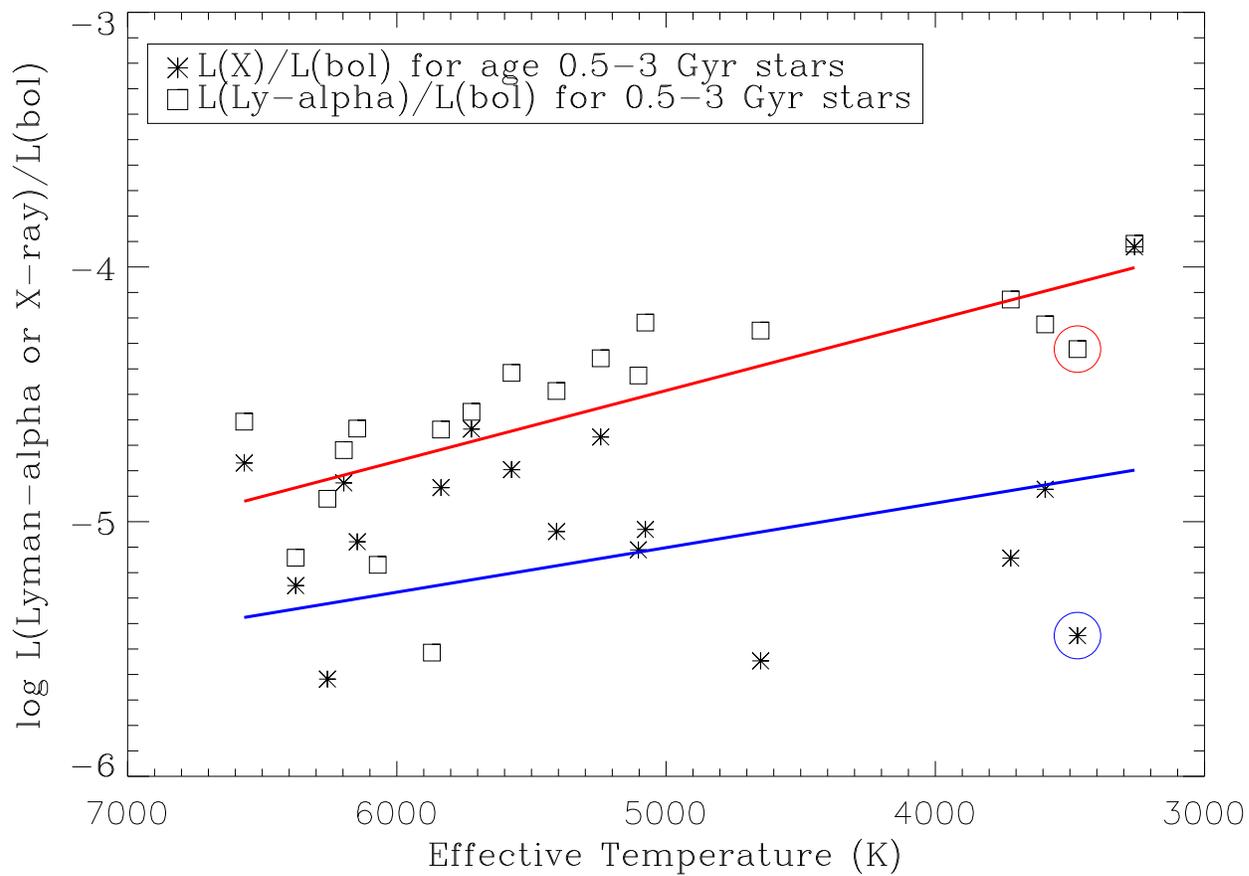}
\end{center}
\caption{Plot of L(X)/L(bol) and L(Ly-$\alpha$)/L(bol) vs. effective
  temperature for stars
  with ages between 0.5 to 3 Gyr. The solid lines are least-squares linear
  fits to the L(Ly-$\alpha$)/L(bol) data (red) and the L(X)/L(bol)
  data (blue). The circuled symbols are for GJ~667C.}
\end{figure}

\clearpage
\begin{figure}[t]
\begin{center}
\includegraphics[angle=0,scale=1.0]{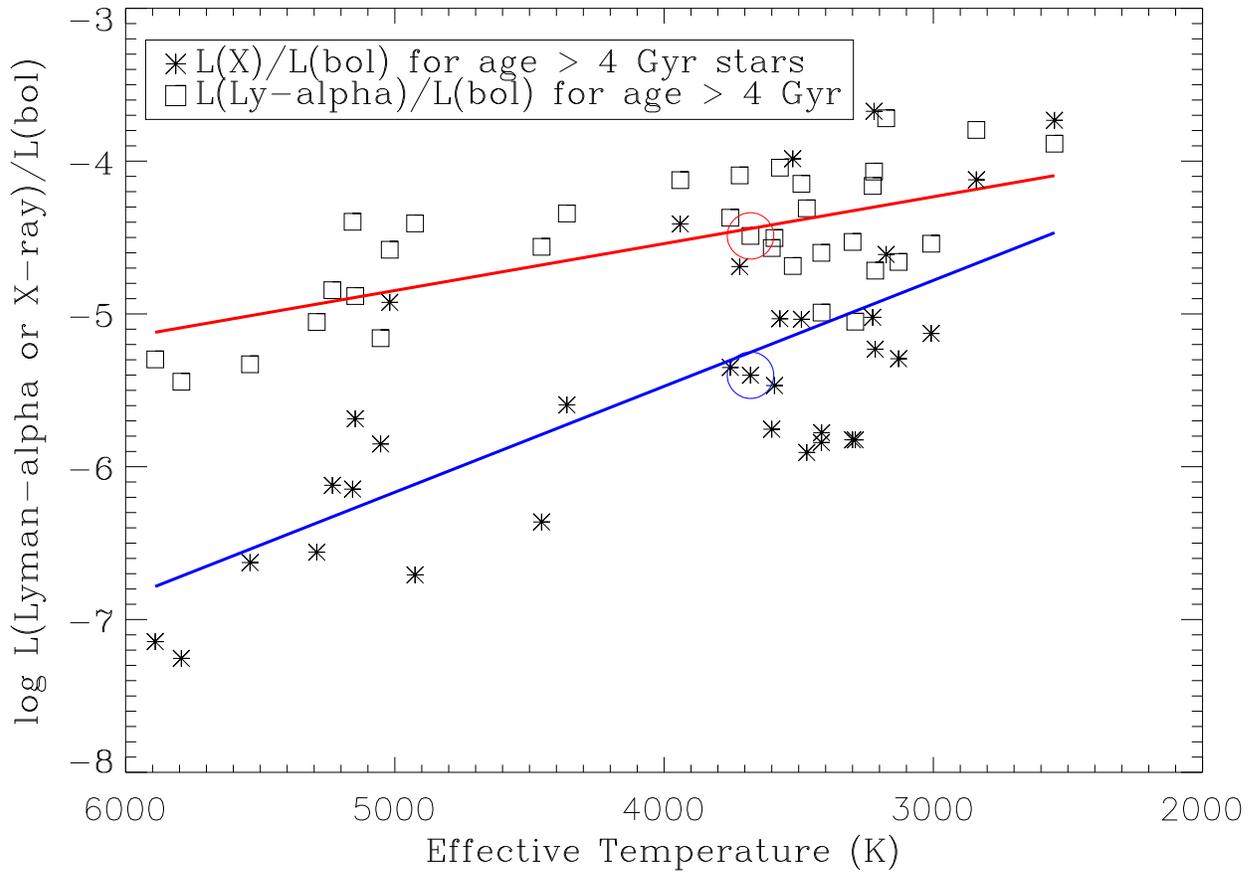}
\end{center}
\caption{Plot of L(X)/L(bol) and L(Ly-$\alpha$)/L(bol) vs. effective
  temperature for stars older
  than 4~Gyr. The solid lines are least-squares linear fits to the
  L(Ly-$\alpha$)/L(bol) data (red) and the L(X)/L(bol)
  data (blue). The circuled symbols are for GJ~176.}
\end{figure}

\clearpage
\begin{figure}[t]
\begin{center}
\includegraphics[angle=0,scale=1.0]{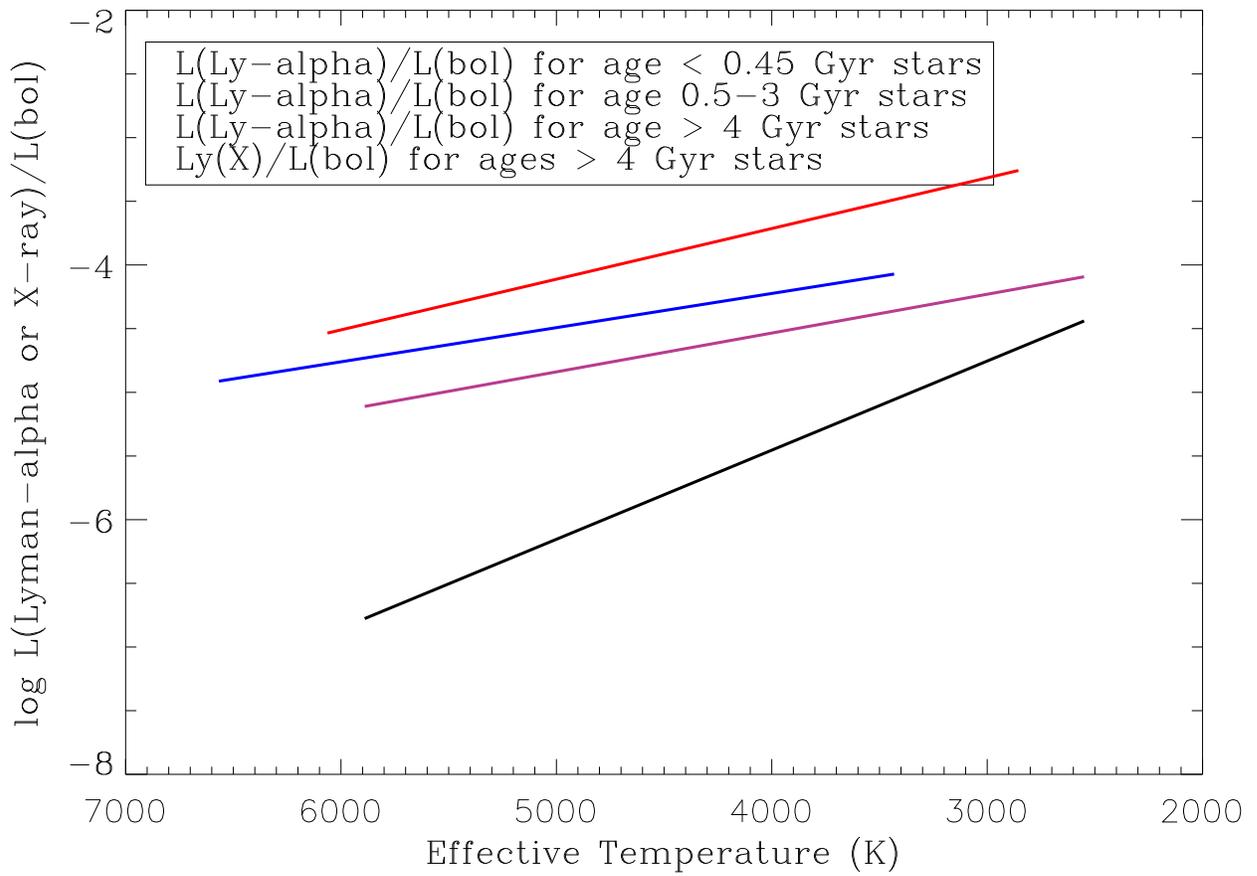}
\end{center}
\caption{Plots of the least-squares fits to the L(Ly-$\alpha$)/L(bol) data
  for young stars ($<0.45$~Gyr, red top line)), middle age stars (0.5--3~Gyr,
  blue second line) and
  older stars ($>4$~Gyr, plum third line) vs effective temperature.
  For comparison, the least-squares fit to the L(X)/L(bol) data is shown
  for the older stars (black line).}
\end{figure}

\clearpage
\begin{figure}[t]
\begin{center}
\includegraphics[angle=0,scale=1.0]{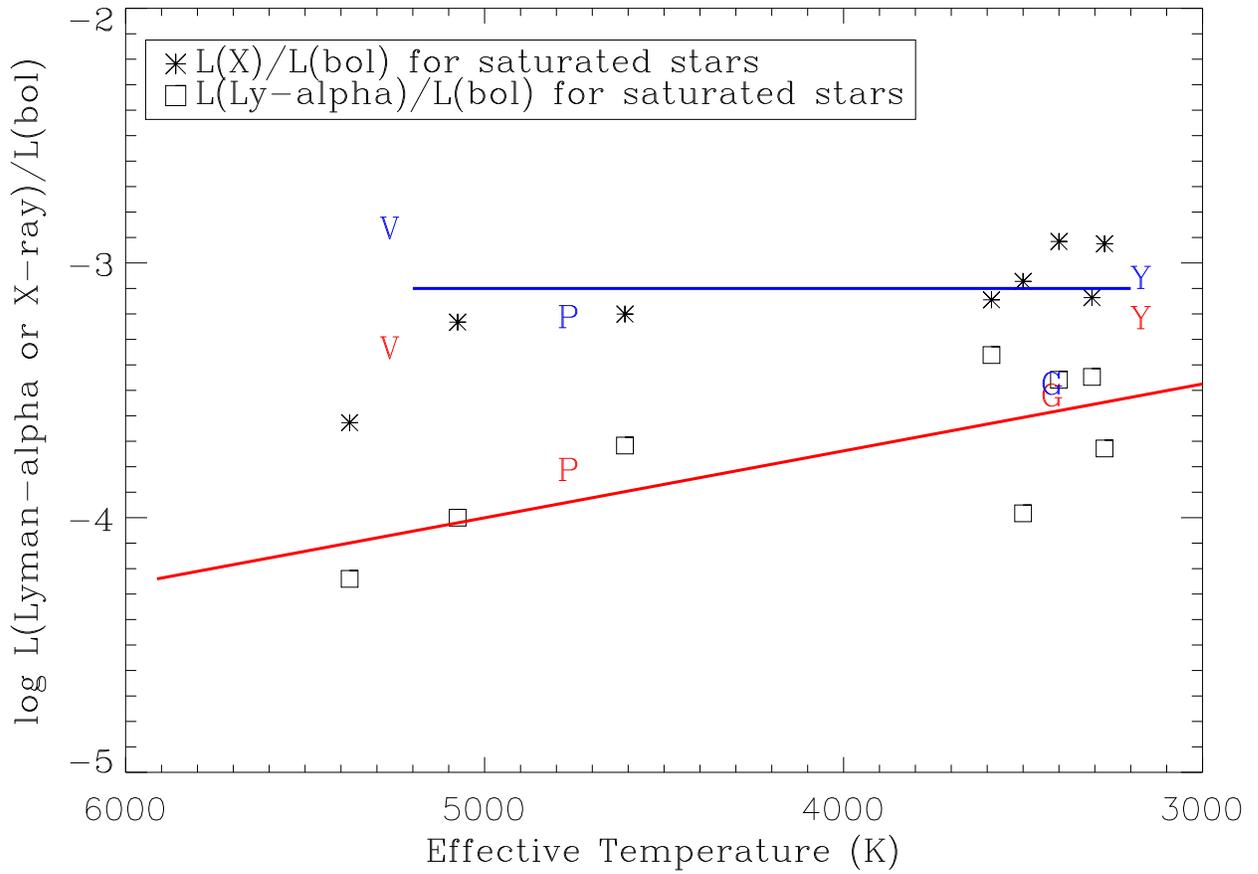}
\end{center}
\caption{Plot of L(X)/L(bol) and L(Ly-$\alpha$)/L(bol) vs.
  effective temperature for stars younger
  than 100~Myr (black symbols). The solid blue line is the X-ray
  saturation level of $10^{-3.1}$. The solid red line is a least-squares
  linear fit to the L(Ly-$\alpha$)/L(bol) data. Also included are
  L(X)/L(bol) (blue) and L(Ly-$\alpha$)/L(bol) (red) data for the
  150~Myr K2~V star PW~And and the field age late-M star YZ CMi
  (P and Y symbols). The V and G
  symbols are for the spectroscopic binaries V471~Tau and GJ~644B.}
\end{figure}

\begin{longrotatetable}
\begin{deluxetable}{lccccccccccc}
\tablewidth{0pt}
\tabletypesize{\footnotesize}
\tablecaption{Stellar Parameters, fluxes (erg~cm$^{-2}$~s$^{-1}$ at 1~au),
  and luminosity ratios
\label{tab:param}}
\tablehead{\multicolumn{2}{c}{Star names (Age group)} & Age(Gyr) &
  $T_{\rm eff}$ & Sp. Type & $d$(pc) & $L_{\rm bol}$ & $f$(Ly-$\alpha$) &
  $f$(X) & R(Ly-$\alpha)$ & R(X) &  Ref.}
\startdata
 SAO 93981(M) & HD 28568 & $1.16\pm 0.82$(S) & 6567 & F2 V & 45.21 & 34.144 & 122.6 & 84.3 & 2.47 & 1.70 & 10,23\\
SAO 76609(M) & HD 28033 & $0.63\pm 0.05$(S) & 6376 & F8 V & 48.38 & 33.983 & 24.7 & 19.2 & 0.722 & 0.561 & 1\\
$\chi$ Her(O) & HD 142373 & $6.85^{+0.42}_{-0.52}$(S) & 5890 & F8 V & 15.83 & 34.083 & 21.7 & 0.309 &0.504&0.00717&1\\
 V376 Peg(M) & HD 209458 & $3.83^{+0.98}_{-0.70}$(S) & 6071 & F9 V & 48.36 & 33.814 & 15.7 & 0.308 & 0.677 & 0.0133 & 1,24\\
HR 4657(M) & HD 106516 & $1.8\pm 0.5$(S) & 6258 & F9 V & 22.35 & 33.802 & 27.8 & 5.43 & 1.23 & 0.241 & 1\\
$\zeta$ Dor(M) & HD 33262 & $0.68\pm 0.47$(S) & 6147 & F9 V & 11.63 & 33.750 & 46.5 & 16.7 & 2.32 & 0.834 &1\\
V993 Tau(M) & HD 28205 & $0.63\pm 0.05$(S) & 6197 & G0 V & 47.78 & 33.913 & 55.5 & 41.4 & 1.91 & 1.42 & 1\\
$\chi^1$ Ori(Y) & HD 39587 & $0.3\pm 0.1$(S) & 5898 & G0 V & 8.840 & 33.602 & 41.6 & 37.3 & 2.92 & 2.62& 1\\
HR 6748(Y) & HD 165185 & $0.44\pm 0.19$(S) & 5932 & G0 V & 17.20 & 33.612 & 48.9 & 53.5 & 3.36 & 3.67 & 1\\
$\pi$~Men(M) & HD 39091 & 3(Y) & 5870 & G0 V & 18.28 & 33.695 & 5.39 & 0.952 & 0.306 & 0.048& 17,6\\
HR 4345(Y) & HD 97334 & $0.45\pm 0.02$(S) & 5906 & G2 V & 22.66 & 33.613 & 42.8 & 39.9 & 2.93 & 2.73 & 1\\
$\alpha$~Cen~A(O) & HD 128620 & $5.3\pm 0.3$(S) & 5793 & G2 V & 1.324 & 33.771 & 7.54 & 0.117 & 0.360 & 0.00557 & 1\\
Quiet Sun(O) & ... & & 5780 & G2 V & ... & 33.586 & 5.95 & 0.224 & 0.436 & 0.0164 & 2\\
Active Sun(O) & ... & & 5780 & G2 V & ... & 33.586 & 9.15 & 2.85 & 0.670 & 0.209 & 2\\
HR 2882(Y) & HD 59967 & $0.35\pm 0.07$(S) & 5830 & G2 V & 21.77 & 33.537 & 55.9 & 41.9 & 4.56 & 3.42 & 1\\
$\kappa^1$ Cet(M) & HD 20630 & $0.6\pm 0.2$(S) & 5723 & G4 V & 9.146 & 33.494 & 30.0 & 25.6 & 2.70 &2.31&1\\
SAO 136111(M) & HD 73350 & $0.51\pm 0.14$(S) & 5836 & G5 V & 24.34 & 33.602 & 32.8 & 19.3 & 2.30 & 1.36 & 1\\
SAO 158720(M) & HD 128987 & $0.62\pm 0.07$(S) & 5574 & G6 V & 23.76 & 33.401 & 34.4 & 14.3 & 3.84 & 1.60 & 1\\
HR 2225(Y) & HD 43162 & $0.32\pm 0.04$(S) & 5651 & G6.5 V & 16.73 & 33.444 & 41.0 & 48.1 & 4.14 & 4.86 & 1\\
$\xi$ Boo A(Y) & HD 131156A & $0.2\pm 0.1$(S) & 5483 & G7 V & 6.733 & 33.365 & 35.3 & 28.3 & 4.28 &3.43& 1\\
61 Vir(O) & HD 115617 & $9.41^{+1.31}_{-3.15}$(S) & 5538 & G7 V & 8.506 & 33.499 & 5.26 & 0.265 & 0.468 & 0.0236 & 1\\
SAO 254993(Y) & HD 203244 & $0.33\pm 0.08$(S) & 5480 & G8 V & 20.81 & 33.371 & 43.8 & 20.2 & 5.24 &2.42 & 1\\
HR 8(Y) & HD 166 & $0.3\pm 0.1$(S) & 5327 & G8 V & 13.78 & 33.368 & 37.9 & 33.0 & 4.56 & 3.94 & 1\\
$\tau$ Cet(O) & HD 10700 & $5.6\pm 1.2$(S) & 5290 & G8.5 V & 3.603 & 33.255 & 5.66 & 0.176 & 0.886 & 0.0276 & 1\\
SAO 28753(Y) & HD 116956 & $0.33\pm 0.10$(S) & 5308 & G9 V & 21.66 & 33.285 & 33.0 & 24.7 & 4.81 & 3.60 &1\\
HR 1925(M) & HD 37394 & $0.5\pm 0.1$(S) & 5243 & K0 V & 12.28 & 33.274 & 29.3 & 14.4 & 4.38 & 2.15 & 1\\
40 Eri A(O) & HD 26965 & $11.76^{+1.92}_{-5.19}$(S) & 5147 & K0.5 V & 5.036 & 33.195 & 7.33 & 1.15 & 1.31 & 0.206 & 1\\
$\alpha$~Cen~B(O) & HD 128621 & $5.3\pm 0.3$(S) & 5232 & K1 V & 1.255 & 33.297 & 10.1 & 0.533 & 1.43 & 0.0756 & 1\\
DX Leo(Y) & HD 82443 & $0.25\pm 0.05$(S) & 5315 & K1 V & 18.08 & 33.256 & 31.1 & 59.7 & 4.85 & 9.30 & 1\\
70 Oph A(M) & HD 165341 & $1.3\pm 0.3$(S) & 5407 & K1 V & 5.122 & 33.308 & 23.6 & 6.62 & 3.26 & 0.915 & 1\\
GJ 3651(O) & HD 97658 & $9.7\pm 2.8$(S) & 5157 & K1 V & 21.57 & 33.101 & 18.01 & 0.321 & 4.01 & 0.0714 & 5,24\\
Kepler-444(O) & HIP 94931 & $11.23^{+0.91}_{-0.99}$(S) & 5053 & K1 V & 36.48 & 33.099 & 3.10 & 0.635 & 0.694 & 0.141 & 10,24\\
EP Eri(Y) & HD 17925 & $0.2\pm 0.1$(S)  & 5167 & K1.5 V & 10.36 & 33.185 & 27.6 & 32.9 & 5.07 & 6.04 & 1\\
$\epsilon$~Eri(M) & HD 22049 & $0.5\pm 0.1$(S) & 5077 & K2 V & 3.203 & 33.092 & 26.6 & 4.10 & 6.05 & 0.932 & 4,5\\
36 Oph A(M) & HD 155886 & $1.7\pm 0.4$(S) & 5103 & K2 V & 5.959 & 33.130 & 18.0 & 3.72 & 3.75 & 0.775 & 1\\
LQ Hya(Y) & HD 82558 & $0.07^{+0.03}_{-0.02}$(S) & 5376 & K2 V & 18.29 & 33.461 & 59.1 & 243.0 & 5.75 & 23.6 & 1\\
V368 Cep(Y) & HD 220140 & $0.09^{+0.06}_{-0.04}$(S) & 5075 & K2 V & 18.96 & 33.120 & 46.9 & 275.0 & 10.0 & 58.6 & 1\\
PW And(Y) & HD 1405 & $0.15^{+0.05}_{-0.02}$(S) & 4796 & K2 V & 28.34 & 32.976 & 47.1 & 187.0 & 14.0 & 55.7 & 1\\
GJ 4130(O) & HD 189733 & $6.4^{+4.8}_{-4.2}$(S) & 5019 & K2 V & 19.77 & 33.102 & 11.8 & 5.34 & 2.63 & 1.19 & 2,19\\
GJ 2046(O) & HD 40307 & $6.9\pm 4.0$(S) & 4925 & K2.5 V & 12.94 & 33.010 & 14.2 & 0.0712 & 3.91  & 0.0196 & 4,5\\
V471 Tau(SB) & ... & $0.63\pm 0.05$(S) & 5291 & K2 V & 47.71 & 33.270 & 277.9 & 1175. & 42.0 & 178. & 1,23\\
Speedy Mic(Y) & HD 197890 & $0.03\pm 0.01$(S) & 4609 & K3 V & 66.76 & 33.497 & 214. & 702. & 19.2 & 63.0,& 1\\
$\epsilon$ Ind(M) & HD 209100 & $1.6\pm 0.2$(S) & 4649 & K4 V & 3.639 & 32.936 & 17.3 & 0.871 &  5.63 & 0.284 & 1\\
61 Cyg A(O) & HD 201091 & $6.0\pm 1.0$(S) & 4361 & K5 V & 3.497 & 32.742 & 8.90 & 0.498 & 4.54 & 0.254 & 1\\
GJ 370(O) & HD 85512 & $5.61\pm 0.61$(S) & 4455 & K6 V & 11.28 & 32.823 & 6.50 & 0.103 & 2.75 & 0.0435 & 4,5\\
GJ 338A(O) & HD 79210J & 5(Y) & 3940 & M0 V & 6.33 & 32.491 & 8.28 & 4.28 & 7.51 & 3.88 & 15\\
Ross 1044(O) & GJ 1188 & $>10$(S) & 3754 & M0 V & 37.17 & 31.996 & 1.50 & 0.157 & 4.26 & 0.446 & 10\\
HIP 23309(Y) & CD-57 1054 & $0.0244$(Y) & 3500 & M0 V & 26.9 & 32.651 & 16.58 & 135. & 10.40 & 84.7 & 16,25\\
AU Mic(Y) & HD 197481 & $0.02\pm 0.01$(S) & 3588 & M1 V & 9.725 & 32.445 & 43.0 & 70.9 & 43.5 & 71.7 & 2\\
Kapteyn's(O) & GJ 191 & $11.5^{+0.5}_{-1.5}$(S) & 3570 & M1 VI & 3.91 & 31.676 & 0.347 & 0.157 & 2.06 & 0.930&7,11\\
GJ 410(Y) & HD 95650 & 0.3(M) & 3775 & M1 V & 11.94 & 32.386 & 9.56 & 12.36 &
11.04 & 14.3 & 16,25\\
GJ 49(O)  & HIP 4872 & 5(M) & 3175 & M1.5 V & 9.86 & 32.180 & 10.27 & 1.32 &
19.06 & 2.45 & 16,24\\
GJ 667C(M) & HD 156384C & $>2$(S) & 3472 & M1.5 V & 7.245 & 31.837 & 1.16 & 0.087 & 4.76 & 0.357 & 4,22\\
GJ 205(O) & HD 36395 & 5(Y) & 3719 & M1.5 V & 5.70 & 32.361 & 6.59 & 1.67 & 8.06 & 2.04 & 15\\
GJ 3470(M) & LP~424-4 & 2(Y) & 3592 & M2 V & 29.45 & 32.235 & 3.64 & 0.82 & 5.95 & 1.34 & 18\\  
GJ 15A(O) & HD 1326 & 5(Y) & 3470 & M2 V & 3.56 & 32.015 & 1.81 & 0.0458 & 4.91 & 0.124 & 14,15,20\\
GJ 887(M) & HD 217987 & 2.9(Y) & 3720 & M2 V & 3.28 & 32.148 & 3.73 & 0.390 & 7.45 & 0.779 & 15\\
GJ 832(O) & HD 204961 & 8.4(S) & 3522 & M2 V & 4.965 & 32.016 & 0.996 & 0.0650 & 2.70 & 0.176 & 4,5\\
GJ 176(O) & HD 285968 & $4.0\pm 0.3$(S) & 3679 & M2 V & 9.473 & 32.113 & 1.49 & 0.183 & 3.24 & 0.397 & 4,5\\
Ross 860(O) & GJ 649 & 5(Y) & 3590 & M2 V & 10.38 & 32.231 & 5.49 & 6.28 & 9.06 & 10.36 & 14,25\\
GJ 588(O) & CD-40 9712 & 5(M) & 3490 & M2.5 V & 5.92 & 32.035 & 2.74 & 0.356 & 7.10 & 0.923 & 15\\
Ross 905(O) & GJ 436 & $4.2\pm 0.3$(S) & 3416 & M3 V & 9.756 & 31.978 & 0.850 & 0.0486 & 2.51 & 0.144 & 4,5\\
GJ 860A$^*$ & HD 239960  & & 3410 & M3 V & 4.01 & 31.782 & 0.750 & 0.935 & 3.48
& 4.34 & 15\\
GJ 581(O) & HO Lib & $4.1\pm 0.3$(S) & 3415 & M3 V & 6.299 & 31.709 & 0.186 & 0.0304 & 1.02 & 0.167 & 4,5\\
GJ 644B(SB)$^*$ & HD 152751 & 5(M) & 3450 & M3.5 V & 6.20 & 32.138 & 13.4 & 14.9 & 27.4 & 30.5 & 15\\
HIP 17695(Y) & G80-21 & 0.1(Y) & 3400 & M3 V & 16.8 & 32.026 & 13.17 & 45.9 & 34.8 & 121.5 & 16,25\\
GJ 674(M) & CD-46 11540 & 0.5(Y) & 3260 & M3 V & 4.55 & 31.654 & 1.98 & 1.91 &  12.34 & 12.0 & 14,24\\
GJ 729(Y) & V1216 Sgr & 0.5(Y) & 3240 & M3.5 V & 2.98 & 32.160 & 1.80 & 3.17 &
34.98 & 61.6 & 14,24\\
Luyten's(O) & GJ 273 & 5(M) & 3290 & M3.5 V & 3.80 & 31.589 & 0.846 & 0.123 & 6.12 & 0.89 & 15\\
GJ 876(O) & IL Aqr & $9.51\pm 0.58$(S) & 3129 & M3.5 V & 4.676 & 31.669 & 0.363 & 0.0846 & 2.19 & 0.509& 4,5\\
GJ 849(O) & BD-05 5(Y) & 5(Y) & 3600 & M3.5 V & 8.80 & 32.107 & 1.43 & 0.156 & 3.14 & 0.34 & 14,20\\
AD Leo(Y) & GJ 388 & 0.025--0.3(S) & 3308 & M3 V & 4.966 & 31.867 & 9.33 & 19.1 & 35.7 & 73.1 & 2\\
GJ 163(O) & L229-91 & 5(Y) & 3225 & M3.5 V & 15.14 & 31.754 & 1.39 & 0.192 & 6.88 & 0.95 &14,20\\
Barnard's(O) & GJ 699 & 10(Y) & 3300 & M4 V & 1.83 & 31.123 & 0.14 & 0.00688 & 2.96 & 0.15 & 11,14,20,22\\
YZ CMi(Y) & GJ 285 & 5(M) & 3200 & M4 V & 5.99 & 31.671 & 9.189 & 13.21 &55.2 & 79.4 & 15\\
GJ 1132(O) & L320-124 & $>5$(S) & 3216 & M4 V & 12.62 & 31.240 & 0.11 & 0.0363 & 1.92 & 0.588 & 14,12,24\\
GJ 1214(O) & G139-21 & 5--10(S) & 3008 & M4 V & 14.65 & 31.065 & 0.119 & 0.0308 & 2.88 & 0.745 & 5,24\\
EV Lac(Y) & GJ 873 & 0.025--0.3(S) & 3273 & M5 V & 5.050 & 31.663 & 3.07 & 19.5 & 18.7 & 119.0 & 1\\
LHS 2686(O) & G177-25 & 5(Y) & 3220 & M5 V & 12.19 & 31.118 & 0.39 & 0.967 & 8.54 & 21.18 &14,20\\ 
Prox Cen(O) & GJ 551 & $5.3\pm 0.3$(S) & 2840 & M5.5 & 1.301 & 30.726 & 0.301 & 0.142 & 16.0 & 7.55 & 2\\
Trappist-1(O) & ... & $7.6\pm 2.2$(S) & 2550 & M7.5 V & 12.43 & 30.301 & 0.092 & 0.131 & 13.0 & 18.5 &
14,8,9\\ 
 \hline
\enddata
\tablerefs{$^*$ The value of $f(X)$ refers to one-half of the total 
X-ray flux from the binary system as the X-ray observations could not
resolve the emission from each star.
R(Ly-$\alpha$)=$10^5$L(Ly-$\alpha$)/L(bol). R(X)=$10^5$L(X)/L(bol).\\
Age group: Y=$\leq 450$~Myr, M=0.5--3~Gy, O=$\geq 4$~Gyr., SB=Spectroscopic
Binary (treated a a young star).\\
Age(Gyr): S = \cite{Schneider2019}, Y = Youngblood (in prep), M = Melbourne
(submitted to ApJ).\\
(1) \cite{Wood2005}; (2) \cite{Linsky2013};
(3) \cite{Youngblood2017}; (4) \cite{Loyd2016}; (5) \cite{Youngblood2016};
(6) \cite{King2019}; (7) \cite{Guinan2016}; 
(8) \cite{Wheatley2017}; (9) \cite{Bourrier2017}; (10) \cite{Schneider2019};
(11) Youngblood high RV program; (12) \cite{Waalkes2019}, (13) \cite{Saur2018};
(14) Mega-Muscles program. (Wilson et al. 2020);
(15) Wood et al. program GO-15326; 
(16) FUMES II paper \cite{Youngblood2020}; (17) \cite{GarciaMunoz2020};
(18) \cite{Bourrier2018}; (19) \cite{Sans-Forcada2011};
(20) Brown et al. (in prep); (21) \cite{Singh1999};
(22) France et al. (2020); (23) {\em ROSAT} Hyades Catalog;
(24) {\em XMM} Seredipitous Source catalog; (25) {\em XMM} Slew Catalog;
(26)\cite{Malo2014}.}
\end{deluxetable}
\end{longrotatetable}
\clearpage

\begin{deluxetable}{clccc}
\tablewidth{0pt}
\tablecaption{Fit Parameters for Figures 2 and 4 and the Pearson correlation coefficients}
\tablehead{Equation & Spectral Type & $\alpha$ & $\beta$ & $r_{\rm Pearson}$}
\startdata
    $\log F_x = \alpha \log F_{Ly\alpha} + \beta$ 
    & F2~V to G9~V &   $2.32\pm 0.26$ & $-2.41\pm 0.38$ & $0.842\pm 0.061$\\
    & K0~V to K7~V &   $2.34\pm 0.31$ & $-2.34\pm 0.45$ & $0.802\pm 0.08$\\
    & M0~V to M2.5~V & $1.76\pm 0.24$ & $-1.00\pm 0.18$ & $0.815\pm 0.086$\\
    & M3~V to M7.5~V & $1.42\pm 0.17$ & $-0.25\pm 0.12$ & $0.832\pm 0.072$\\
$\log L_X/L_{bol} = \alpha \log L_{Ly\alpha}/L_{bol} + \beta$ 
    & F2~V to G9~V &   $2.28\pm 0.20$ & $-0.85\pm 0.09$ & $0.888\pm 0.046$\\
    & K0~V to K7~V &   $2.27\pm 0.42$ & $-1.36\pm 0.33$ & $0.693\pm 0.114$\\ 
    & M0~V to M2.5~V & $1.97\pm 0.44$ & $-1.39\pm 0.39$ & $0.655\pm 0.138$\\
    & M3~V to M7.5~V & $1.89\pm 0.18$ & $-1.06\pm 0.19$ & $0.88\pm 0.052$\\
\hline
\enddata
\end{deluxetable}


\begin{thebibliography}{99}
\expandafter\ifx\csname natexlab\endcsname\relax\def\natexlab#1{#1}\fi

\bibitem[Ayres(2014)]{Ayres2014} Ayres, T.~R.\ 2014, \aj, 147, 59

\bibitem[Baraffe \& Chabrier(2018)]{Baraffe2018} Baraffe, I. \&
Chabrier, G.\ 2018, \aap, 619, A177

\bibitem[Berger et al.(2010)]{Berger2010} Berger, E., Basri, G.,
  Fleming, T.~A.\ 2010, \apj, 709, 332

\bibitem[Bourrier et al.(2017)]{Bourrier2017} Bourrier, V.,
  Ehrenreich, D., Wheatley, P.~J., et al.\ 2017, \aap, 599, L3

\bibitem[Bourrier et al.(2018)]{Bourrier2018} Bourrier, V., Lecavelier
  des Etangs, A., Ehrenreich, D.\ 2018, \aap, 620, A147


\bibitem[Brun \& Browning(2017)]{Brun2017} Brun, A.~S. \& Browning,
  M.~K.\ 2017, Living Rev. Solar Phys., 14, 4

\bibitem[Cameron, Dikpati \& Brandenburg(2017)]{Cameron2017} Cameron, R.~H.,
  Dikpati, M., \& Brandenburg, A.\ 2017, \ssr, 210, 367
  
\bibitem[Chabrier \& Baraffe(2000)]{Chabrier2000} Chabrier, G.
\& Baraffe, I.\ 2000, \araa, 38, 337

\bibitem[Charbonneau(2014)]{Charbonneau2014} Charbonneau, P.\ 2014,
\araa, 52, 251

\bibitem[Claire et al.(2012)]{Claire2012} Claire, M.~W., Sheets, J.,
  Cohen, M.\ 2012, \apj, 757, 95

\bibitem[Cranmer \& Winebarger(2019)]{Cranmer2019} Cranmer, S.~R.,
  Winebarger, A.~R.\ 2019, \araa, 57, 157

\bibitem[Dobler(2005)]{Dobler2005} Dobler, W.\ 2005, Astr. Nachrichten,
  326, 254

\bibitem[Drake, Laming \& Widing(1997)]{Drake1997} Drake, J.~J., Laming, J.~M.,
  Widing, K.~J.\ 1997, \apj, 478, 403  
  
\bibitem[Durney, De Young, \& Roxburgh(1993)]{Durney1993} Durney, B.~R., 
De Young, D.~S., \& Roxburgh, I.~W.\ 1993, \solphys, 145, 207

\bibitem[Feldman \& Laming(2000)]{Feldman2000} Feldman, U.,
  Laming, K.~G.\ 2000, Phys. Scripta, 61, 222

\bibitem[Fleming, Schmitt, \& Giampapa(1995)]{Fleming1995} Fleming,
  T.~A., Schmitt, J.~H.~M.~M., \& Giampapa, M.~S.\ 1995, \apj, 450, 401

\bibitem[Fontenla et al.(2016)]{Fontenla2016} Fontenla, J., Linsky, J.~L.,
Witbrod, J., et al.\ 2016, \apj, 830, 154

\bibitem[France et al.(2018)]{France2018} France, K., Arulanantham, N.,
Fossati, L., et.al.\ 2018, \apjs, 239, 16

\bibitem[France et al.(2012)]{France2012} France, K., Linsky, J.~L.,
Tian, F., Froning, C.~S., \& Roberge, A.\ 2012, \apj, 750, L32

\bibitem[France et al.(2016)]{France2016} France, K., Loyd, R.~O.~P.,
Youngblood, A., et al.\ 2016, \apj, 820, 89

\bibitem[Froning et al.(2019)]{Froning2019} Froning, C.~S, Kowalski, A.,
  France, K., et al.\ 2019, \apjl, 871, L26

\bibitem[Garcia Munoz et al. (2020)]{GarciaMunoz2020} Garcia Munoz A., 
  Youngblood, A., Fossati, L., Gandolfi, D., Rauer,, H. \ 2020, \apj,
  888, L21

\bibitem[Guinan, Engle, \& Durbin(2016)]{Guinan2016} Guinan, E.~F., 
Engle, S.~G., \& Durbin, A.\ 2016, \apj, 821, 81

\bibitem[G\"udel(2004)]{Gudel2004} G\"udel, M.\ 2004,
  Astron. Astrophys. Rev. 12, 71

\bibitem[G\"udel et al.(2001)]{Gudel2001} G\"udel, M., Audard, M., Briggs, K.,
  et al.\ 2001, \aap, 365, L336

\bibitem[Hussain et al.(2006)]{Hussain2006} Hussain, G.~A.~J.,
  Allende Prieto, C., Saar, S.~H., Still, M.\ 2006, \mnras, 367, 1699
  
\bibitem[Jackson, Davis, \& Wheatley(2012)]{Jackson2012} Jackson, A.~P.,
  Davis, T.~A., Wheatley, P.~J.\ 2012, \mnras, 422, 2024 

\bibitem[Jao et al.(2018)]{Jao2018} Jao, W.-C., Henry, T.~J., Gies,
  D.~R., \& Hambly, N.~C.\ 2008, \apj, 861, L11

\bibitem[Hawley \& Johns-Krull(2003)]{Hawley2003} Hawley, S., \&
  Johns-Krull, C.~M.\ 2003, \apj, 588, L109

\bibitem[Kaltenegger(2017)]{Kaltenegger2017} Kaltenegger, L.\ 2017,
  \araa, 55, 433

\bibitem[King et al.(2019)]{King2019} King, G.~W., Wheatley, P.~J.,
    Bourrier, V., Ehrenreich, D.\ 2019, \mnras, 484, L49

\bibitem[Laming(2015)]{Laming2015} Laming, J.~M.\ 2015, Living Rev. Solar
    Phys., 12, 2
    
\bibitem[Linsky(2017)]{Linsky2017} Linsky, J.~L.\ 2017,
  Ann. Rev. Astron. Astrophys. 55, 197

\bibitem[Linsky, France, \& Ayres(2013)]{Linsky2013} Linsky, J.~L., France, K.,
\& Ayres, T.\ 2013, \apj, 766, 69

\bibitem[Loyd \& France(2014)]{Loyd2014} Loyd, R.~O.~P. \& France, K.\ 2014,
  \apjs, 211, 9

\bibitem[Loyd et al.(2016)]{Loyd2016} Loyd, R.~O.~P., France, K.,
Youngblood, A., et al.\ 2016, \apj, 824, 102

\bibitem[Loyd et al.(2018)]{Loyd2018} Loyd, R.~O.~P., France, K.,
Youngblood, A., et al.\ 2018, \apj, 867, 71

\bibitem[Malo et al.(2014)]{Malo2014} Malo, L., Artigau, E., Doyon, R.,
  et al.\ 2014, \apj, 788, 81 

\bibitem[Marino et al.(2002)]{Marino2002} Marino, A., Micela, G., Peres, G.,
  \& Sciortino, S.\ 2002, \aap, 383, 210

\bibitem[Mazeh et al.(2001)]{Mazeh2001} Mazeh, T., Latham, D.~W., Goldberg, E.,
  et al.\ 2001, \mnras, 325, 343
  
\bibitem[Melbourne et al.(2020)]{Melbourne2020} Melbourne, K., Youngblood, A.,
  France, K., et al.\ 2020, submitted to \apj

\bibitem[Miles \& Shkolnik(2017)]{Miles2017} Miles, B.~E.\&
  Shkolnik, E.~L.\ 2017, \aj, 154, 67
  
\bibitem[Morin et al.(2010)]{Morin2010} Morin, J., Donati, J.-F., Petit, P.,
  Delfosse, X., Forveille, T., \& Jardine, M.~M.\ 2010, \mnras, 407, 2269

\bibitem[Newton et al.(2017)]{Newton2017} Newton, E.~R., Irwin, J.,
  Charbonneau, D., Berlind, P., Calkins, M.~L., \& Mink, J.\ 2017,
  \apj, 834, 85

\bibitem[Oranje(1986)]{Oranje1986} Oranje, B.~J.\ 1986, \aap, 154, 185

\bibitem[Pizzolato et al.(2003)]{Pizzolato2003}  Pizzolato, N., Maggio, A.,
  Micela, G., Sciortino, S., Ventura, P.\ 2003, \aap, 397, 147
  
\bibitem[Rebull et al.(2017)]{Rebull2017} Rebull, L.~M., Stauffer, J.~R.,
Hillenbrand, L.~A., et al.\ 2017, \apj, 839, 92

\bibitem[Reiners(2012)]{Reiners2012} Reiners, A.\ 2012, Living
  Rev. Sol. Phys., 9, 1

\bibitem[Reiners \& Mohanty(2012)]{Reiners-Mohanty2012} Reiners, A., Mohanty,
  S.\ 2012, \apj, 746, 43

\bibitem[Ribas et al.(2005)]{Ribas2005} Ribas, I., Guinan, E.~F., 
G\"udel, M., \& Audard, M.\ 2005, \apj, 622, 680

\bibitem[Richey-Yowell et al.(2019)]{Richey-Yowell2019} Richey-Yowell, T.,
Shkolnik, E.~L., Schneider, A.~C., Osby, E., Barman, T.,
Meadows, V.~S.\ 2019, \apj, 872, 17

\bibitem[Robrade \& Schmitt(2009)]{Robrade2009} Robrade, J. \&
Schmitt, J.~H.~M.~M.\ 2009, \aap, 497, 511

\bibitem[Rutten et al.(1989)]{Rutten1989} Rutten, R.~G.~M., Schrijver,
  C.~J., Zwaan, C., Duncan, D.~K., Mewe, R.\ 1989, \aap, 219, 239

\bibitem[Sans-Forcada et al.(2011)]{Sans-Forcada2011} Sans-Forcada, J.,
Micela, G., Ribas, I.\ 2011, \aap, 532, A6

\bibitem[Saur, et al.(2018)]{Saur2018} Saur, J., Fischer, C.,
  Wennmacher, A., et al.\ 2018, \apj, 859, 74

\bibitem[Saxton et al.(2008)]{Saxton2008} Saxton, R.~D., Read, A.~M.,
  Esquej, P., et al.\ 2008, \aap, 480, 611
  
\bibitem[Schmidt et al.(2015)]{Schmidt2015} Schmidt, S.~J., Hawley, S.~L.,
West, A.~A., et al.\ 2015, \aj, 149, 158

\bibitem[Schneider \& Shhkolnik(2018)]{Schneider2018} Schneider, A.~C. \&
  Shkolnik, E.~L.\ 2018, \aj, 155, 122

\bibitem[Schneider et al.(2019)]{Schneider2019} Schneider, A.~C.,
  Shkolnik, E.~L., Barman, T.~S., Loyd, R.~P.\ 2019,
  \apj, 886, 19 

\bibitem[Schrijver \& Rutten(1987)]{Schrijver1987} Schrijver, C.~J.,
  Rutten, R.~G.~M.\ 1987, \aap, 177, 143

\bibitem[Shields, Ballard, \& Asher(2016)]{Shields2016} Shields, A.~L.,
  Ballard, S., \& Asher, J.\ 2-16, Physics Reports, 663, 1S38

\bibitem[Shulyak et al.(2015)]{Shulyak2015} Shulyak, D., Reiners, A.,
  Seemann, U., Kochukhov, O., \& Piskunov, N.\ 2015, \aap, 563, A35
  
\bibitem[Singh et al.(1999)]{Singh1999}  Singh, K.~P., Drake, S.~A.,
  Gotthelf, E.~V., White, N.~E.\ 1999, \apj, 512, 874
  
\bibitem[Stelzer et al.(2012)]{Stelzer2012} Stelzer, B., Alcal\'a, J.,
Biazzo, K., et al.\ 2012, \aap, 537, 94

\bibitem[Stelzer et al.(2013)]{Stelzer2013} Stelzer, B., Marino, A.,
  Micela, G., L\'opez-Santiago, Liefke, C.\ 2013, \mnras, 431, 2063

\bibitem[van Saders \& Pinsonneault(2012)]{vanSaders2012} van Saders, J.~L. \&
Pinsonneauldt, M.~H.\ 2012, \apj, 751, 98

\bibitem[von Steiger et al.(1995)]{vonSteiger1995} von Steiger, R.,
  Wimmer-Schweingruber, R.~F., Geiss, J., Gloeckler, G.\ 1995,
  Adv. Space Res., 15, 3

\bibitem[Walkowicz \& Hawley(2009)]{Walkowicz2009} Walkowicz, L.~M. \&
Hawley, S.~L.\ 2009, \aj, 137, 3297

\bibitem[Waalkes et al.(2019)]{Waalkes2019} Waalkes, W.~C.,
  Berta-Thompson, Z., Bourrier, V., et al.\ 2019, \aj, 158, 50

\bibitem[Wandel(2018)]{Wandel2018} Wandel, A.\ 2018, \apj, 856, 165

\bibitem[Wheatley(1998)]{Wheatley1998} Wheatley, P.~J.\ 1998, \mnras, 297, 1145
  
\bibitem[Wheatley et al.(2017)]{Wheatley2017} Wheatley, P., Loudon, T.,
Bourrier, V., Ehrenreich, D., \& Gillon, M.\ 2017, \mnras, 465, L74  

\bibitem[Wilson et al.(2020)]{Wilson2020} Wilson, D.~J., Froning, C.~S.,
  Duvvuri, G.~M., et. al\ 2020, submitted to \apj

\bibitem[Wood et al.(2018)]{Wood2018} Wood, B.~E., Laming, J.~M.,
  Warren, H.~P., Poppenhaeger, K.\ 2018, \apj, 862, 66

\bibitem[Wood et al.(2005)]{Wood2005} Wood, B.~E., Redfield, S., Linsky, J.~L.,
M\"uller, H.-R., \& Zank, G.~P.\ 2005, \apjs, 159, 118

\bibitem[Wright \& Drake(2016)]{Wright2016} Wright, N.~J. \& 
Drake, J.~J.\ 2016, Nature, 535, 526

\bibitem[Youngblood et al.(2016)]{Youngblood2016} Youngblood, A., France, K.,
Loyd, R.~T.~O.~P., et al.\ 2016, \apj, 824, 101

\bibitem[Youngblood et al.(2017)]{Youngblood2017} Youngblood, A.,
France, K., Loyd, R.~T.~O.~P., et al.\ 2017, \apj, 843, 31

\bibitem[Youngblood et al.(2020)]{Youngblood2020} Youngblood, A.,
  Pineda, J.~S., France, K.\ 2020, submitted to \apj


\end{thebibliography}
\end{document}